\documentclass[12pt,a4paper]{article}
\usepackage[utf8]{inputenc}
\usepackage[english]{babel}

\usepackage{graphicx}
\usepackage{amssymb}
\usepackage{amsmath}
\usepackage[colorlinks=true,linkcolor={blue},citecolor={green},urlcolor={red}]{hyperref}
\usepackage{cite}

\newcommand{\bm}[1]{{\boldsymbol{#1}}}
\renewcommand{\vec}[1]{{\bm{#1}}}

\newcommand{\pdiff}[2]{{\dfrac{\partial{#1}}{\partial{#2}}}}

\newcommand{\spdiff}[2]{{{\partial{#1}}/{\partial{#2}}}}

\begin{document}

\title%
{%
Full 3-D MHD calculations of accretion flow Structure in magnetic cataclysmic 
variable stars with strong and complex magnetic fields
}%

\author%
{%
 Zhilkin A.G.$^{1,2}$\thanks{E-mail: zhilkin@inasan.ru}, %
 Bisikalo D.V.$^{1}$, %
 Mason P.A.$^{3,4}$\\ %
\textit{\small $^{1}$ Institute of Astronomy, Russian Academy of Sciences, Moscow, Russia}\\%
\textit{\small $^{2}$ Chelyabinsk State University, Chelyabinsk, Russia}\\%
\textit{\small $^{3}$ University of Texas at El Paso, El Paso, USA}\\%
\textit{\small $^{4}$ New Mexico State University, Las Cruces, USA}\\%
}%
\date{}

\maketitle%

\begin{abstract}
\scriptsize
We performed 3-D MHD calculations of stream accretion in cataclysmic variable
stars for which the white dwarf primary star possesses a strong and complex
magnetic field. These calculations are motivated by observations of polars;
cataclysmic variables containing white dwarfs with magnetic fields sufficiently
strong to prevent the formation of an accretion disk. So, an accretion stream
flows from the $L_1$ point and impacts directly onto one or more spots on the
surface of the white dwarf. Observations indicate that the white dwarf, in some
binaries, possesses a complex (non-dipolar) magnetic field. We perform 
simulations of ten polars, where the only variable is the azimuthal angle of 
the secondary with respect to the white dwarf. These calculations are also 
applicable to asynchronous polars, where the spin period of the white dwarf is 
a few percent different from the orbital period. Our results are equivalent to 
calculating the structure of one asynchronous polar at ten different spin-orbit 
beat phases. Our models have an aligned dipole plus quadrupole magnetic field 
centered on the white dwarf primary. We find that for a sufficiently strong
quadrupole component an accretion spot occurs near the magnetic equator for
slightly less than half of our simulations while a polar accretion zone is 
active for most of the rest of the simulations. For two configurations, 
accretion at the dominant polar region and at an equatorial zone occurs
simultaneously. Most polar studies assume that the magnetic field structure is 
dipolar, especially for single pole accretors. We demonstrate that for orbital 
parameters and magnetic field strengths typical of polars, accretion flow 
patterns are widely variable in the presence of a complex magnetic field. We 
suggest that it might be difficult to observationally determine if the field is 
a pure dipole or if it is complex for many polars, but there will be 
indications for some systems. Specifically, a complex magnetic field should be 
considered if the there is an accretion zone near the white dwarf's spin 
equator (assumed to be in the orbital plane) or if there are two or more 
accretion regions that cannot be fit by a dipole magnetic field. For 
asynchronous polars, magnetic field constraints are expected to be 
substantially stronger, with clearer indicators of complex field geometry due
to changes in accretion flow structure as a function of azimuthal angle. These 
indicators become more apparent in asynchronous polars since each azimuthal
angle corresponds to a different spin-orbit beat phase. 
\end{abstract}

\section{Introduction}

Magnetic cataclysmic variables (mCVs) are close binary systems consisting of a
low mass late type star (secondary) undergoing mass transfer onto a white dwarf 
(primary) that possesses a surface magnetic field strong enough to constrain 
plasma flow. The envelope of the secondary star fills its Roche lobe and 
overflow occurs through the vicinity of the inner Lagrangian point $L_1$ 
\cite{Warner1995}. The vast majority of cataclysmic variables are non-magnetic 
as magnetic effects become apparent only for ($>10^7$~G) fields. One can pick 
out two main types of magnetic cataclysmic variables (mCVs), namely 
intermediate polars and polars.

In polars (AM Her type systems) the induction of the magnetic field on the 
white dwarf's surface is highest among mCVs with an observed range 7--230~MG. 
These systems have relatively short orbital periods (1--5~hours) and the 
rotation of the component stars is in, or is nearly in, synchronism with the 
binary orbit. It is clear that accretion disks do not form in polars and the
matter flowing from the secondary forms a collimated stream moving along
magnetic lines toward one or more hot spots on the white dwarf. The strong
magnetic field leads to a high degree of polarization, hence the name polar, of
the optical/IR emission coming from the vicinity of the accretion region(s). 
Some polars rotate asynchronously, with spin and orbital periods differing by a 
few percent, yet they display most or all of the other characteristics of 
synchronized polars.

The presence of complex magnetic fields in white dwarfs~--- defined as a 
magnetic field which cannot be adequately modelled as a centered or offset 
dipole, dates at least back to \cite{Schmidt1986}. They modelled PG 1031+234 
with an oblique centered dipole with the addition of a high magnetic field 
spot on the magnetic equator. Wu and Wickramasinghe \cite{Wu1992} suggested 
that in the presence of a complex magnetic field, a band-like region near the 
intersection of the magnetic equator and the orbital plane should be the 
primary accretion zone in polars. Detailed studies of a number of polars 
revealed the need to involve complex magnetic fields (e.g. 
\cite{Schwope1995}). However, these cases could also be explained in terms of 
highly offset magnetic dipoles. Strong observational evidence exists for a 
complex magnetic field in the asynchronous polar BY Cam \cite{Mason1995, 
Silber1997, Mason1998}. Recently the existence of complex magnetic fields in 
white dwarfs was confirmed by Zeeman tomography of several polars 
\cite{Euchner2002, Euchner2005, Euchner2006, Beuermann2007}. The aim of the 
current work is to calculate stream trajectories and to locate accretion zones 
in polars using full 3-D magneto-hydrodynamic simulations in order to 
determine various accretion flow modes in polars with complex magnetic fields. 
These calculations are performed as a function of azimuthal angle of the 
secondary. In particular, to investigate equatorial accretion zones which only 
occur for complex magnetic fields. 

Following Occam's razor, the elimination of pure dipoles, centered or 
otherwise, leads to the simplest type of complex field. Such a field, to a
first approximation, may be represented as a superposition of aligned dipole
and quadrupole modes. The accretion characteristics in this case may strongly
differ from the case of a purely dipole field. In particular such a
configuration may have several magnetic poles, that, in turn, can lead to
formation of several accretion zones. The defining observational characteristic
of such a field is the possibility of accretion on or near the magnetic
equator. However, given such a field, the accretion flow pattern is expected 
to vary as a function of the azimuthal angle of the secondary in relation to 
the magnetic field axis. In this paper we calculate accretion flow patterns at 
ten positions covering the full range of azimuthal angles. 

\begin{figure}[t]
\centering
\includegraphics[width=0.95\textwidth]{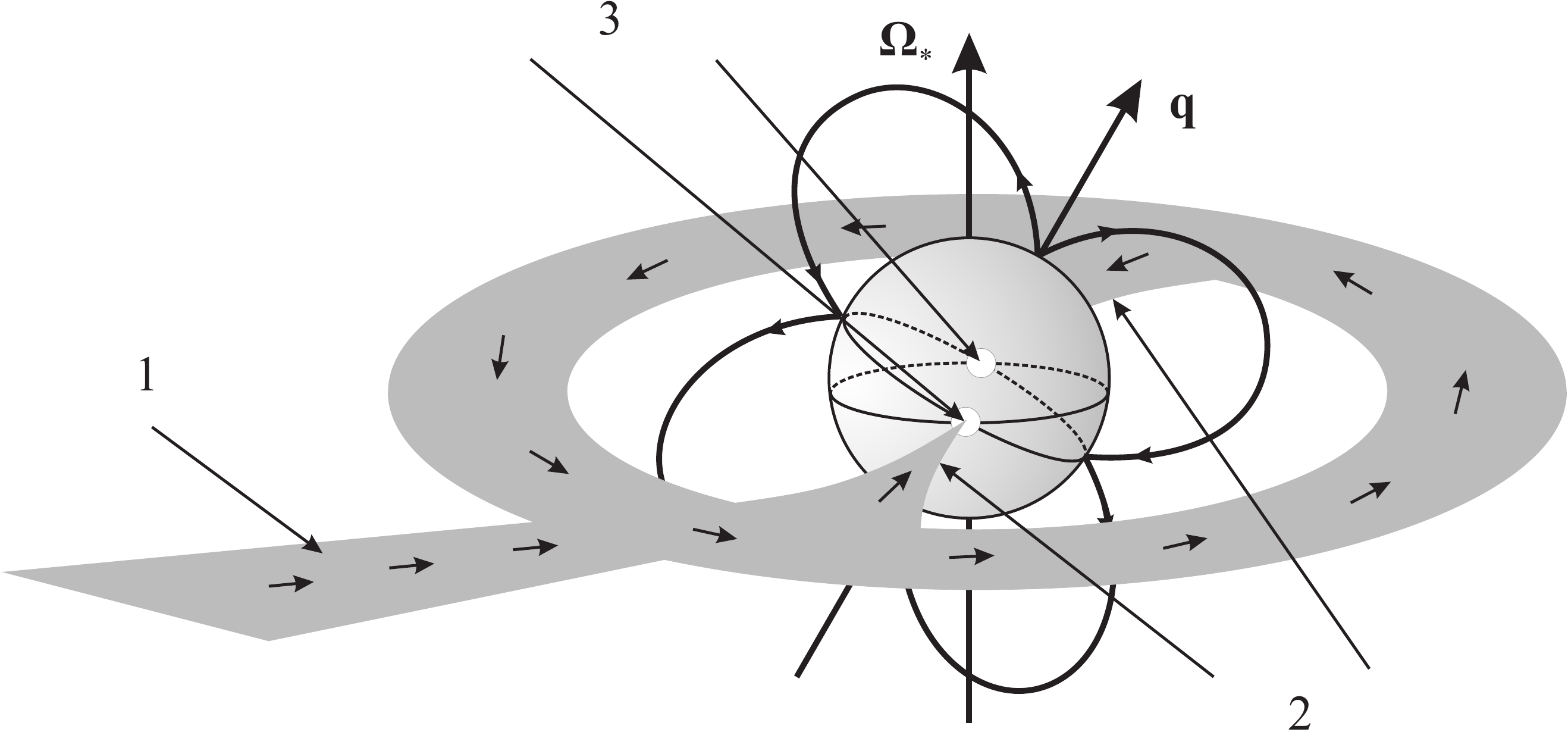}%
\caption{The scheme explaining how matter may be accreted by a compact gravitating 
object that has a strong pure quadrupole magnetic field. The axes of the 
primary star's rotation and symmetry of the quadrupole magnetic field are 
shown by vectors $\vec{\Omega}_{*}$ and $\vec{q}$ respectively. The lines 
of the geographic and magnetic equator are shown. The numbers are corresponded 
to: 1~--- accretion stream, 2~--- accretion columns, 3~--- quadrupole hot 
spots.}
\label{fg-scheme}
\end{figure}

Intermediate polars have accretion disks disrupted by the magnetic field at the
inner edge of the disk. Figure \ref{fg-scheme} schematically demonstrates the
main features of disk accretion in a close binary system where the accretor
possesses a strong purely quadrupole magnetic field. The axis of symmetry of
the field (denoted by the vector $\vec{q}$) is inclined with respect to the
rotation axis of the star (denoted by the vector of the angular velocity of the
star' proper rotation $\vec{\Omega}_{*}$). The matter of the accretion stream
forms a ring in the immediate vicinity of the white dwarf's surface, since the
magnetic field does not allow it to fall directly onto the surface. The
accreted matter has significant angular momentum. Thus, the plane of symmetry
of the ring almost coincides with the orbital plane of the binary system. In
the case of the quadrupole magnetic field, matter can get onto the accretor's
surface either in regions of the magnetic poles or along the magnetic equator.
However, consideration of the conservation law of angular momentum, one
can see that the most effective regime is when matter is accreted onto the
points of intersection of the magnetic (red line) and geographic (blue line)
equators of the star. Exactly at these points (yellow circles) accretion hot
spots will form. We must emphasize that at these points the magnetic field has
no preferred direction since the magnetic lines coming from the northern and
southern magnetic poles have different directions. Hence the radiation
originating in these zones will not be polarized in pure quadrupole fields, in
contrast with the radiation originating in hot spots on the magnetic poles. We
suggest that models of this type may be applicable to some intermediate polars,
but more detailed studies covering more of the parameter space are required.

Asynchronous polars, also known as BY Cam stars (coined by Patterson
\cite{Patterson1994}) offer observations that strongly constrain magnetic
field configurations. This results from the fact that periodic (or
quasi-periodic) variations in accretion flow geometry result as the accretion
flow evolves. This is because the $L_1$ point originates from all azimuthal
angles, with respect to the magnetic field, over the spin-orbit beat cycle. By
the present time several ($\sim 4$) asynchronous polars have been confirmed.
The most known representative of this subtype is the BY Cam system itself, in
which the period of the proper rotation of the white dwarf differs from the
orbital period by approximately $1\%$ (see e.g. \cite{Mason1998}). There is 
significant evidence for a complex magnetic field from polarimetry observations 
\cite{Mason1998, Piirola1994} and as such the accretion flow cannot be 
adequately modelled by an offset dipole field \cite{Mason1998}.

The asynchronous rotation of the white dwarf in BY Cam and the complex 
structure of its magnetic field result in light curves that evolve as a 
function spin-orbit beat phase \cite{Silber1997, Mason1998, Schwarz2005, 
Pavlenko2007a, Andronov2008}. These studies generally support the complex 
field model of Mason et al. \cite{Mason1995}. Specifically, in BY Cam there is 
a $\sim$ 2 minute difference between the spin and orbital periods allowing for 
the sampling of the magnetic field structure at all phases of the 14.1 day 
spin-orbit beat-cycle. This is a bit like dropping iron filings all around a 
magnet in the classic physics demonstration, to examine field structure. 
Synchronized polars sample accretion flow from a single azimuthal angle. So, 
by calculating the accretion flow at a variety of azimuthal angles we 
simultaneously create models for 10 polars and a single asynchronous polar, 
with the spin-orbit characteristics of BY Cam. This allows for a study of 
polars as function of azimuthal angle. It also provides an initial attempt to 
model the pole switching process in BY Cam. It is hoped that further 
exploration of poorly constrained parameters such as accretion rate and mass 
ratio are needed before a proper model of BY Cam can be expected.

3D numerical simulations of mass transfer process in semi-detached binary 
systems without taking into account the magnetic field of the accretor star 
was performed in papers \cite{Bisikalo2000, Bisikalo2001, Bisikalo2003, 
Bisikalo2004a, Bisikalo2004b, Bisikalo2005, Sytov2007, Bisikalo2008, 
Sytov2009a, Sytov2009b}. In these works a self-consistent numerical model for 
gas-dynamics in close binary systems without magnetic field was first 
developed and the basic characteristics of the flow structure was obtained.

Recently authors \cite{ZB2009, ZhilkinMM2010} developed a 3D parallel 
numerical code for simulations of the flow structure in close binary systems 
that takes into account the proper magnetic field of the accretor. In recent 
works \cite{ZB2009, ZBASR2010, ZBMFG2010}, the code was utilized to 
investigate the flow structure in a close binary system (SS Cyg was taken as 
an example) with a relatively weak magnetic field ($10^5$~G on the surface) 
of the pure dipole type. In \cite{ZBSMF2010} a modification of the code was 
introduced. This modification allows for simulations of systems with strong 
dipole magnetic fields ($10^7$--$10^8$~G).

Pioneering 3-D MHD simulations of the disk-like accretion onto a star with
complex geometry of the magnetic field have been performed \cite{Long2007,
Long2008}. Results of these simulations showed that in the case of the
disk-like accretion onto a star with a purely quadrupole magnetic field the
matter of the disk is accreted in the vicinity of the quadrupole magnetic belt.
If the dipole component is taken into account the belt shifts toward the
southern magnetic pole since in the northern hemisphere the total magnetic
field is stronger than in the southern hemisphere. Romanova et al. 
\cite{Romanova2011} considered more complex configurations of the magnetic 
field where an octupole component exists. These models were used to model the 
disk-like accretion onto young stars of the T Tau type.

In this paper we describe a modification of the numerical model designed to
simulate systems where accretors have a complex magnetic field structure. We
simulate the flow structure in polars with a complex magnetic field including 
aligned dipole and quadrupole components. These calculations are performed at 
ten different azimuthal angles thereby calculating ten separate polar models. 
Putting these together yields a single asynchronous polar model. We are able to 
see how in polars, single spot accretion may be misleading; and how accretion 
in asynchronous polars provide powerful constraints on magnetic field 
configurations.

The paper is composed as follows. In the second section we discuss the
configuration of the magnetic field, describe the governing equations and the
numerical method used in the frame of the modified model. In the third section
results of the 3-D simulations of the flow structure are presented and possible 
observational indicators of flow features are discussed. In the final section 
the main results of the work are briefly discussed and summarized.

\section{Description of the model}

\subsection{Magnetic field}

To describe plasma flow in a close binary system it is convenient to use a
non-inertial reference frame that rotates with the system having angular
velocity $\vec{\Omega}$ with respect to its center of mass. In this frame we
choose a Cartesian coordinate system ($x$, $y$, $z$). The origin of the
coordinate system is placed at the center of the accretor and the center of
the donor is placed at the distance $A$ from the accretor on the $x$-axis. The
$z$ axis is directed along the rotation axis of the binary. The secondary 
transfers matter through the inner Lagrangian point $L_1$ since in this point 
the pressure gradient is not balanced by gravity.

Let us consider a close binary system with the accretor rotating synchronously,
i.e. its rotational period is equal to the orbital one. In the absence of
currents in the envelope of the binary the magnetic field will be the proper
field of the accretor $\vec{B}_{*}$. The magnetic field of the accretor in
the region of the magnetosphere may be rather strong. Thus the magnetic field
in plasma $\vec{B}$ is conveniently treated as a superposition of the
accretor's field and the field induced by currents in the plasma: $\vec{B} =
\vec{B}_{*} + \vec{b}$.

This implies that in the envelope the proper magnetic field of the accretor
must satisfy the condition $\nabla \times \vec{B}_{*} = 0$ since it is
governed by currents inside the accreting star. Thus, in the envelope this
field can be described with the scalar potential: $\vec{B}_{*} =
-\nabla\varphi$. The complex magnetic field of the accretor is taken to be
a superposition of the multipole components of the field. Further we assume
that every multipole component of the field is axisymmetric. In general the
axes of symmetry of different multipole components may not be coincident. In
our current model, we take into account only the dipole and quadrupole 
components of the magnetic field of the accretor.

The magnetic induction corresponding to the dipole component of the field is
\begin{equation}\label{eq2}
 \vec{B}_d = \frac{\mu}{r^3}
 \left[ 3 (\vec{d} \cdot \vec{n}) \vec{n}  - \vec{d} \right],
\end{equation}
where $\mu$ is the magnetic moment of the accretor, $\vec{d}$ is a unit
vector that determines the axis of symmetry of the dipole field, $\vec{n} =
\vec{r}/r$. The vector of the magnetic moment is $\vec{\mu} = \mu
\vec{d}$. Let us denote the value of the magnetic induction on the magnetic
pole of the dipole field of the star as $B_{d,\text{a}}$. Then, using
\eqref{eq2} the magnetic moment is $\mu = B_{d,\text{a}} R^3_{\text{a}}/2$, 
where $R_{\text{a}}$ is the accretor's radius.

The potential corresponding to the quadrupole component of the field is
characterized by a tensor of the quadrupole moment $D_{ik}$. If the
distribution of currents inside the accretor is axisymmetric then the diagonal
components of the quadrupole moment are $D_{11} = D_{22} = -D/2$, $D_{33} = D$
and all the non-diagonal components are zero. In this case the induction of the
corresponding magnetic field can be described as
\begin{equation}\label{eq5}
 \vec{B}_q = \frac{3D}{4r^4}
 \left[ 5 (\vec{q} \cdot \vec{n})^2 \vec{n} - \vec{n} - 
 2 (\vec{q} \cdot \vec{n}) \vec{q} \right],
\end{equation}
where $\vec{q}$ is a unit vector that determines the axis of symmetry of the
quadrupole field. Let us denote the value of the magnetic induction on the
magnetic pole of the quadrupole field of the star as $B_{q,\text{a}}$. Then,
using \eqref{eq5} the quadrupole moment is $D = 2 B_{q,\text{a}} R^4_{\text{a}}
 / 3$. 

In our model the total magnetic field of the accretor is $\vec{B}_{*} =
\vec{B}_d + \vec{B}_q$. We assume that the axes of symmetry of the dipole
and quadrupole fields are coincident ($\vec{d} = \vec{q}$). In this case in
calculations it is convenient to set $B_{d,\text{a}} = \alpha_{d} 
B_{\text{a}}$, $B_{q,\text{a}} = \alpha_{q} B_{\text{a}}$, where 
$B_{\text{a}}$ is the total induction of the field on the magnetic pole of 
the accretor and the coefficients $\alpha_{d}$ and $\alpha_{q}$ satisfy the 
relation $\alpha_{d} + \alpha_{q} = 1$ and determine the relative 
contributions of the dipole and quadrupole components into the total value of 
the magnetic field. In the plane of the magnetic equator $\vec{d} \cdot 
\vec{n} = 0$, $\vec{q} \cdot \vec{n} = 0$. Thus the dipole and quadrupole 
field inductions are:
\begin{equation}\label{eq6}
 B_d = \frac{B_{d,\text{a}}}{2} \frac{R^3_{\text{a}}}{r^3}, \quad 
 B_q = \frac{B_{q,\text{a}}}{2} \frac{R^4_{\text{a}}}{r^4}.
\end{equation}

The ratio of these values is $B_d / B_q = (\alpha_d / \alpha_q) (R_{\text{a}}
 / r)$. This relation demonstrates that if $\alpha_q > \alpha_d$ then in the
vicinity of the magnetic equator a certain region will exist where the
quadrupole field predominates over the dipole field.

\begin{figure}[t]
\centering
\includegraphics[width=0.75\textwidth]{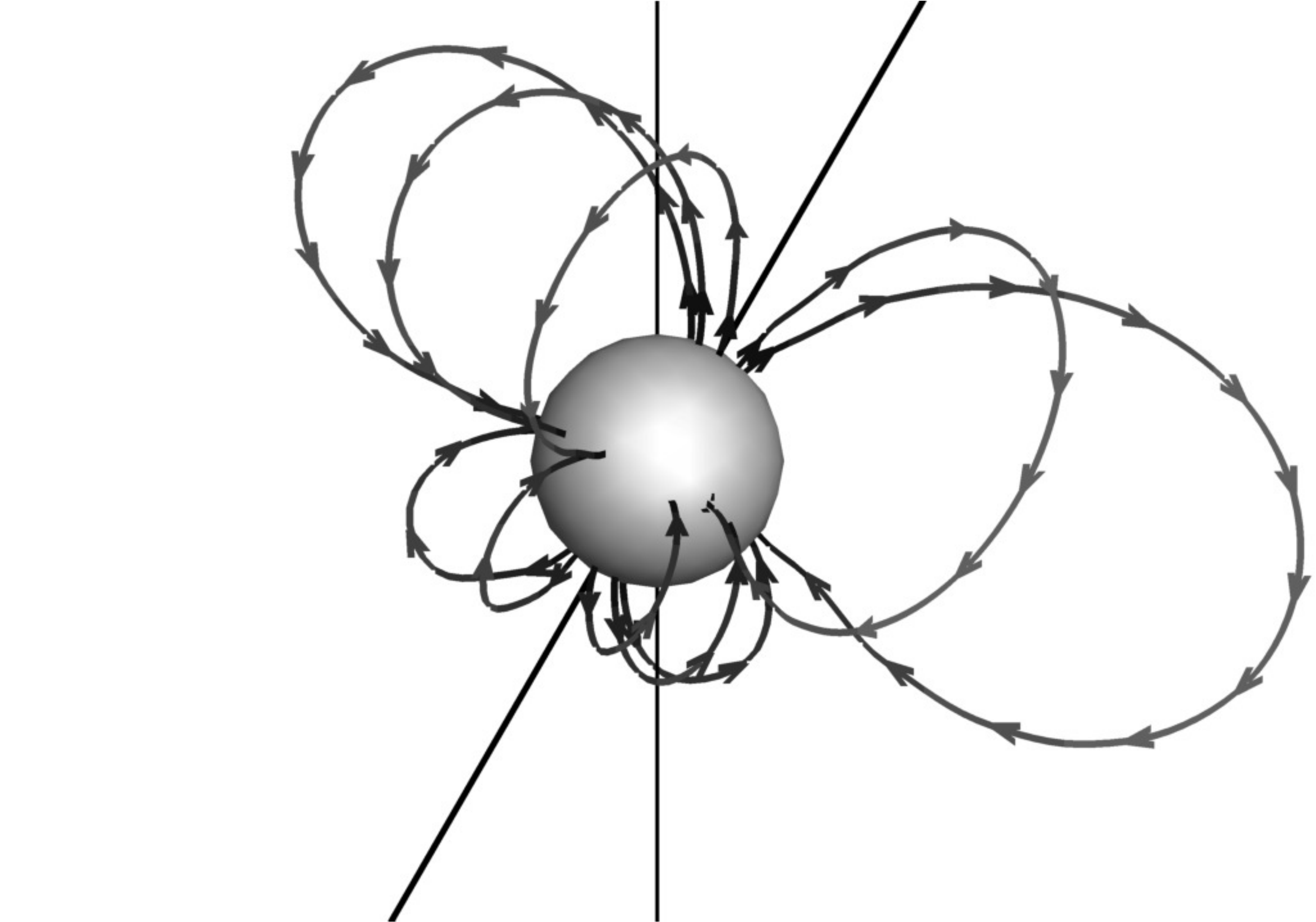}
\caption{Geometry of the accretor's magnetic field. The gray scale of the field lines 
corresponds to the value of the magnetic induction $B_{*}$. The rotation axis 
of the accretor and the magnetic axis are shown.}
\label{fg-mf}
\end{figure}

The geometry of the magnetic field we use in our calculations is shown in Fig.
\ref{fg-mf}. The lines with the arrows are the field lines. The gray scale of
the lines corresponds to the value of the magnetic induction. The vertical 
line is the accretor's rotaiton axis. The inclined line denotes the axis of 
symmetry of the magnetic field. The maximal value of the induction on the 
accretor's surface is achieved on the magnetic poles and along the line 
located near the magnetic equator. For a purely quadrupole field this line 
coincides with the magnetic equator. If the dipole component is also taken 
into account this line slightly shifts southwards.

Indeed, let us denote the scalar product as $\vec{d} \cdot \vec{n} =
\cos\theta$. The normal component of the magnetic field on the accretor's
surface (taking into account \eqref{eq2} and \eqref{eq5} is equal to
\begin{equation}\label{eq6a}
 \vec{n} \cdot \vec{B}_{*} = 
 B_{\text{a}} \left(
 \alpha_d \cos\theta + 
 \frac{3}{2} \alpha_q \cos^2\theta - 
 \frac{3}{2} \alpha_q
 \right).
\end{equation}
Calculating the derivative with respect to $\theta$ we can find values of the
angle $\theta_{*}$ corresponding to maximal values of the magnetic field on 
the accretor's surface. For $\theta_{*}$ we have the equation:
\begin{equation}\label{eq6b}
 \alpha_d \sin\theta_{*} + 3 \alpha_q \cos\theta_{*} \sin\theta_{*} = 0.
\end{equation}

For a purely dipole field we have two solutions: $\theta_{*} = 0$, $\pi$. They
correspond to magnetic poles. For a purely quadrupole field ($\alpha_d = 0$,
$\alpha_q = 1$) we have one more solution: $\theta_{*} = \pi/2$. This solution
corresponds to the magnetic equator. Finally, in the general case we have
\begin{equation}\label{eq6c}
 \cos\theta_{*} = -\frac{\alpha_d}{3\alpha_q}.
\end{equation}
This means that if we regard both the dipole and quadrupole components the line
of the strong field lies a little lower than the magnetic equator ($\theta_{*} 
> \pi/2$).

\subsection{Basic equations}

Plasma dynamics in a strong external magnetic field are characterized by the 
relatively slow average motion of particles along the field lines, their drift 
across the field lines, and propagation of very rapid Alfv\'{e}n and 
magneto-sonic waves against a background of this slow motion. Over the 
characteristic dynamical time scale, the MHD waves can cross the flow region 
(along a column-like stream, for example) many times. Therefore, we can
investigate the average flow pattern, considering the influence of the fast 
pulsations analogous to the MHD wave turbulence. To describe the slow motion 
of plasma itself, it is necessary to separate out rapidly propagating 
fluctuations and apply a well defined procedure for averaging over the 
ensemble of wave pulsations. Such a model for description of MHD flows in 
polars was developed in our recent work \cite{ZBSMF2010}. 

Plasma flow that occurs in a close binary system due to mass transfer can be
described using the following system of \textit{modified MHD} equations taking
into account the strong magnetic field $\vec{B}_{*}$ of the compact object:
\begin{equation}\label{eq7}
 \pdiff{\rho}{t} + \nabla \cdot \left( \rho\vec{v} \right) = 0,
\end{equation}
\begin{equation}\label{eq8}
 \pdiff{\vec{v}}{t} + \left(\vec{v} \cdot \nabla \right) \vec{v} =
 -\dfrac{\nabla P}{\rho} -
 \dfrac{\vec{b} \times (\nabla \times \vec{b})}{4\pi\rho} + 
 2 (\vec{v} \times \vec{\Omega}) - \nabla\Phi -
 \dfrac{\vec{v}_{\perp}}{\tau},
\end{equation}
\begin{equation}\label{eq9}
 \pdiff{\vec{b}}{t} =
 \nabla \times \left[
  \vec{v} \times \vec{b} +
  \vec{v} \times \vec{B}_{*} -
  \eta (\nabla \times \vec{b})
 \right],
\end{equation}
\begin{equation}\label{eq10}
 \rho T \left[
 \pdiff{s}{t} + (\vec{v} \cdot \nabla) s
 \right] =
 n^2 \left(\Gamma - \Lambda\right) +
 \dfrac{\eta}{4\pi}(\nabla \times \vec{b})^2.
\end{equation}
Here $\rho$ is the density, $\vec{v}$ is the velocity, $P$ is the pressure,
$s$ denotes the entropy per unit mass, $n = \rho/m_{\text{p}}$ is the
concentration, $m_{\text{p}}$ is the proton mass, $\eta$ is the coefficient of
the magnetic viscosity and $\Phi$ is the Roche potential. In deriving these
equations we assumed the accretor's magnetic field to be potential ($\nabla
\times \vec{B}_{*} = 0$) and stationary ($\spdiff{\vec{B}_{*}}{t} = 0$).

In the entropy equation \eqref{eq10} we take into account effects of radiative
heating and cooling as well as matter heating due to the dissipation of
currents (last term). It should be noted that the functions of radiative
heating and cooling $\Gamma$ and $\Lambda$ depend on the temperature in a
complicated manner (see \cite{Cox1971, Dalgarno1972, Raymond1976,
Spitzer1981}). In our numerical model we use a linear approximation of these
functions in the vicinity of the equilibrium temperature $T = 11230$~K
\cite{Bisikalo2003, ZB2009, ZhilkinMM2010} that corresponds to the effective
temperature of the accretor of $37000$~K. The term $2(\vec{v} \times 
\vec{\Omega})$ in the equation of motion \eqref{eq8} describes the Coriolis 
force. The density, entropy and pressure are connected via the equation of 
state of the ideal gas: $s = c_V \ln(P / \rho^{\gamma})$. Here $c_{V}$ is the 
gas specific heat at constant volume and $\gamma = 5/3$ is the adiabatic index.

The last term in the equation of motion \eqref{eq8} describes the force the
magnetic field on the gas flow. This force influences the plasma velocity
component $\vec{v}_{\perp}$ perpendicular to magnetic field lines. We note
that the particles' motion perpendicular to field lines is mainly caused by 
the gravity of the compact object (gravitational drift, see, e.g., 
\cite{FrankKamenetsky1968, Chen1987, Trubnikov1996}). Due to the Larmor 
character of the particles' motion in the magnetic field their average motion 
in the perpendicular to the field direction decreases. A strong magnetic field 
plays the role of an effective fluid interacting with the plasma. Thus, the 
last term in \eqref{eq8} may be regarded as a force of friction occurring 
between the plasma and magnetic field. This term is equivalent to the friction 
force acting in a plasma containing different types of particles 
\cite{FrankKamenetsky1968, Chen1987}.

In polars no accretion disk forms and the accretion stream looks like a funnel
flow moving from the inner Lagrangian point $L_1$ toward a magnetic pole of 
the accretor. In this case the main effect leading to the currents' 
dissipation will probably be MHD wave turbulence conditioned by the Alfv\'{e}n 
and magneto-sonic waves propagating through the funnel accretion flow. if this 
is the case, the velocity of these waves is much higher than the proper 
velocity of the plasma flow and in some cases may even be relativistic. The 
coefficient of the magnetic field's diffusion $\eta$ concerned with the MHD 
turbulence can be estimated using the relation
\begin{equation}\label{eq11}
 \eta = \dfrac{\tau_w}{3} \langle \delta \vec{v}^2 \rangle,
\end{equation}
where $\tau_w$ is the correlation time of pulsations, $\delta\vec{v}$ is the
velocity fluctuations and the angle brackets mean values averaged over the 
wave pulsations ensemble. This relation may be parametrized as follows
\begin{equation}\label{eq12}
 \eta = \alpha_w \dfrac{B_{*} l_w}{\sqrt{4\pi\rho}},
\end{equation}
where $\alpha_w$ is a dimensionless parameter close to the unity that
determines the efficiency of the wave diffusion and $l_w$ is the specific
spatial scale of the pulsations. This term may be set using the scale of
inhomogeneity of the background magnetic field $l_w = {B_*}/{|\nabla B_{*}|}$.
In the demonstrated calculations $\alpha_{w} = 0.3$. The relaxation time 
is taken to be $\tau = 4\pi\rho\eta/B_{*}^2$.

\subsection{Numerical method}

For mCV simulations we used the 3-D parallel numerical code Nurgush 
\cite{ZB2009, ZhilkinMM2010, ZBASR2010, ZBSMF2010}. The code is based on a
finite-difference Godunov type scheme of a high approximation order. An
original method of unified variables for MHD \cite{Zhilkin2007} allowed us
to use adaptive meshes in the code. In the calculation described below we used
a geometrically adaptive mesh that was concentrated near the accretor's
surface. It allowed us to significantly increase the resolution of the
magnetosphere region. To minimize numerical errors in the finite-difference
scheme only the magnetic field induced by currents in the accretion stream and
outer envelope is calculated \cite{Tanaka1994, Powell1999}. To remove the
divergence of the magnetic field we use an eight-wave method \cite{Powell1999,
Dellar2001}.

The equation describing the diffusion of the magnetic field was solved using 
an implicit locally one-dimensional method with the factorized operator
\cite{Samarsky1989}. It should be noted that this equation in the curvilinear
non-orthogonal coordinate system, conditioned by the adaptive mesh, contains 
mixed spatial derivatives. To solve this problem our method use the 
regularization of the factorized operator. In fact, the regularization is done 
by substitution of the operators-multipliers constituting the initial 
factorized operator with certain equivalent tridiagonal operators. Hence, the 
regularization parameter is determined by the maximal absolute eigenvalue of 
the metric tensor describing the curvilinear coordinate system. The resulting 
system of linear algebraic equations with a tridiagonal matrix is solved 
numerically using the Thomas algorithm.

\section{Results of simulations}

\subsection{Parameters}

In order to explore polar accretion in the presence of a complex field as a 
function of azimuthal angle we consider the properties of the asynchronous 
polar BY Cam. In addition to providing an ensemble of ten synchronized polars,
it is a first step in modelling asynchronous polars. We investigate the flow 
structure in an interacting binary whose parameters correspond to an 
asynchronous polar, with the same orbital and rotational periods as BY Cam 
(see, e.g., \cite{Mason1995}). Other properties of the BY Cam binary are not 
well constrained and so are generally estimated here. A more detailed attempt 
at modelling BY Cam specifically awaits future work, since this requires 
confrontation with observations at each spin-orbit beat phase. Modelling  
an asynchronous polar, in this manner, is significantly more constraining 
than a synchronized polar, since the field structure at each azimuthal angle 
is sampled as an asynchronous polar progresses through its spin-orbit beat 
cycle. 

The donor star (red dwarf) in our model has the mass $M_{\text{d}} = 
0.5~M_{\odot}$ and an effective temperature of $4000$~K. The accretor (white 
dwarf) has the mass $M_{\text{a}} = 1~M_{\odot}$. The orbital period of the 
system is $P_{\text{orb}} = 3.36$~hours and its semi-major axis is $A = 
1.3~R_{\odot}$. The inner Lagrangian point $L_1$ is located at the distance of 
$0.57~A$ from the accretor's center. The rotational period of the accretor in 
BY Cam is $P_{\text{spin}} = 3.32$~hours. This value differs from the value of 
the orbital period by approximately $1\%$. In the reference frame associated 
with the binary the accretor makes complete rotation in $P_{\text{beat}} = 
14.1$~days ($1/P_\text{beat} = 1/P_\text{spin} - 1/P_\text{orb}$) that is 
about 101 orbital periods. However, the actual beating and dynamics of pole 
switching within BY Cam is not modelled and importantly, the spin and beat 
periods are independent of our calculations, since we are assuming synchronous 
rotation for each of ten azimuthal angles. Hence, the beat phase calculations 
are applicable to any asynchronous polar as long as the asynchronism is no 
more than a few percent.

The surface magnetic induction of the white dwarf in BY Cam is estimated from
observations of cyclotron humps as $28$~MG \cite{Schwope1991}. This is a very 
typical magnetic field strength for polars. In our model the axes of the 
dipole and quadrupole field are coincident. The angle of inclination of the 
magnetic field with respect to the axis of the accretor's rotation is chosen 
to be $30^\circ$. In the calculations we used a value of the coefficient 
$\alpha_q = 10 \alpha_d$. This means that on the magnetic equator, the 
quadrupole field component is 10 times larger than the dipole component. With 
that, the radius of the affected zone of the quadrupole field (a distance at 
which values of $B_d$ and $B_q$ become equal (see \eqref{eq6}) may be 
estimated as $r = 10~R_{\text{a}}$. Of course, the presence of a strong 
magnetic field significantly influences the process of the mass transfer in 
mCVs. In addition, the presence of the quadrupole component can noticeably 
complicate the character of the accretion of matter onto the white dwarf. 
Thus, information on the flow structure obtained using 3-D numerical 
simulations, and their confrontation with observations, will promote a deeper 
understanding of the physical processes involving the interaction of matter 
with complex magnetic fields. The results in this paper offer a start to the 
exploration of the vast parameter space, largely unconstrained by 
observations, available to polars by examining accretion flow structure as a 
function of azimuthal angle of the secondary with respect to the magnetic 
field.

The following initial and boundary conditions were utilized in our model, 
based on values thought typical in polars. At the inner Lagrangian point 
$L_1$ we set the gas velocity equal to the local sound speed $c_s = 
7.4~\text{km}/\text{s}$ corresponding to the donor's effective temperature of 
$4000$~K. The density of gas in $L_1$ is $\rho(L_1) = 
1.1~\times~10^{-7}~\text{g}/\text{cm}^{3}$ and the mass transfer rate is 
$\dot{M} = 10^{-9}~M_{\odot}/\text{yr}$. At the other boundaries of the 
computational domain, the following boundary conditions were set up: the 
density was $\rho_{\text{b}} = 10^{-6}~\rho(L_1)$; the temperature was 
$T_{\text{b}} = 11230$~K; the velocity was $\vec{v}_{\text{b}} = 0$ and the 
magnetic field was $\vec{B}_{\text{b}} = \vec{B}_{*}$. The accretor was set 
up as a sphere of radius $0.0125~A$ at whose boundary the free inflow boundary
condition was defined. With that, the plasma inflow vector was made parallel 
to the vector $\vec{B}_{*}$. All the matter having entered the cells occupied 
by the accretor is considered to have fallen onto the white dwarf. The initial
conditions in the entire computational domain were as follows: the density was
$\rho_0 = 10^{-6}~\rho(L_1)$; the temperature was $T_0 = 11230$~K; the 
velocity was $\vec{v}_0 = 0$ and the magnetic field was $\vec{B}_0 = 
\vec{B}_{*}$. The task was solved in the computational domain of the dimension 
of ($-0.56A \le x \le 0.56A$, $-0.56A \le y \le 0.56A$, $-0.28A \le z \le 
0.28A$) in the geometrically adaptive mesh \cite{ZhilkinMM2010} containing 
$128~\times~128~\times~64$ cells. The calculations presented below were 
performed using the computational facility of the Joint Supercomputer Center 
of the Russian Academy of Sciences.

Numerical experiments showed that the solution achieves the quasi-stationary
regime in a specific time approximately equal to the orbital period 
$P_{\text{orb}}$. The criterion testifying that the solution is in the
quasi-stationary regime is the condition of constancy of the entire mass of
matter in the computational domain. The specific time during which the 
solution becomes quasi-stationary is considerably shorter than 
$P_{\text{beat}}$ of our model asynchronous polar,so indeed our results are 
applicable to asynchronous polars with beat periods of several days or more. 
Thus, in this work we performed simulations of the flow structure separately 
for different azimuthal angles of polars, but we may apply these to phases of 
the $P_{\text{beat}}$ period disregarding the proper rotation of the accretor. 
Totally we modelled ten phases corresponding to time moments 
$m P_{\text{beat}}$, where $m$ varied from $0$ to $0.9$ with the step of $0.1$ 
(this corresponds formally to ten different possible polars). The initial 
phase $m = 0$ corresponds to the accretor's position when the northern 
magnetic pole was located at the opposite from the donor side of the accretor 
(i.e. on the positive arm of the $x$ axis).

\subsection{Flow structure}

\begin{figure}[t]
\includegraphics[width=0.45\textwidth]{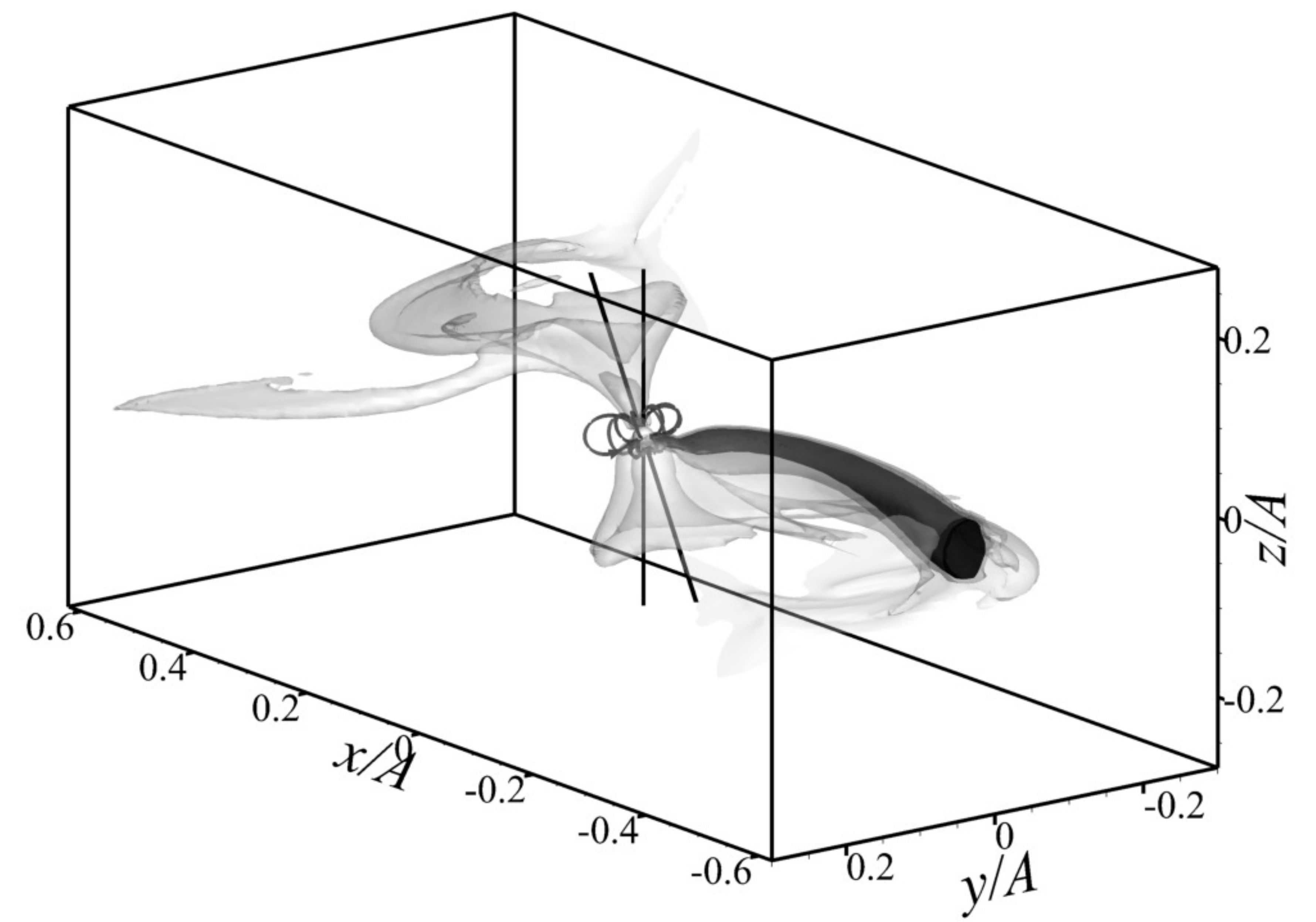}%
\hspace{0.1cm}
\includegraphics[width=0.45\textwidth]{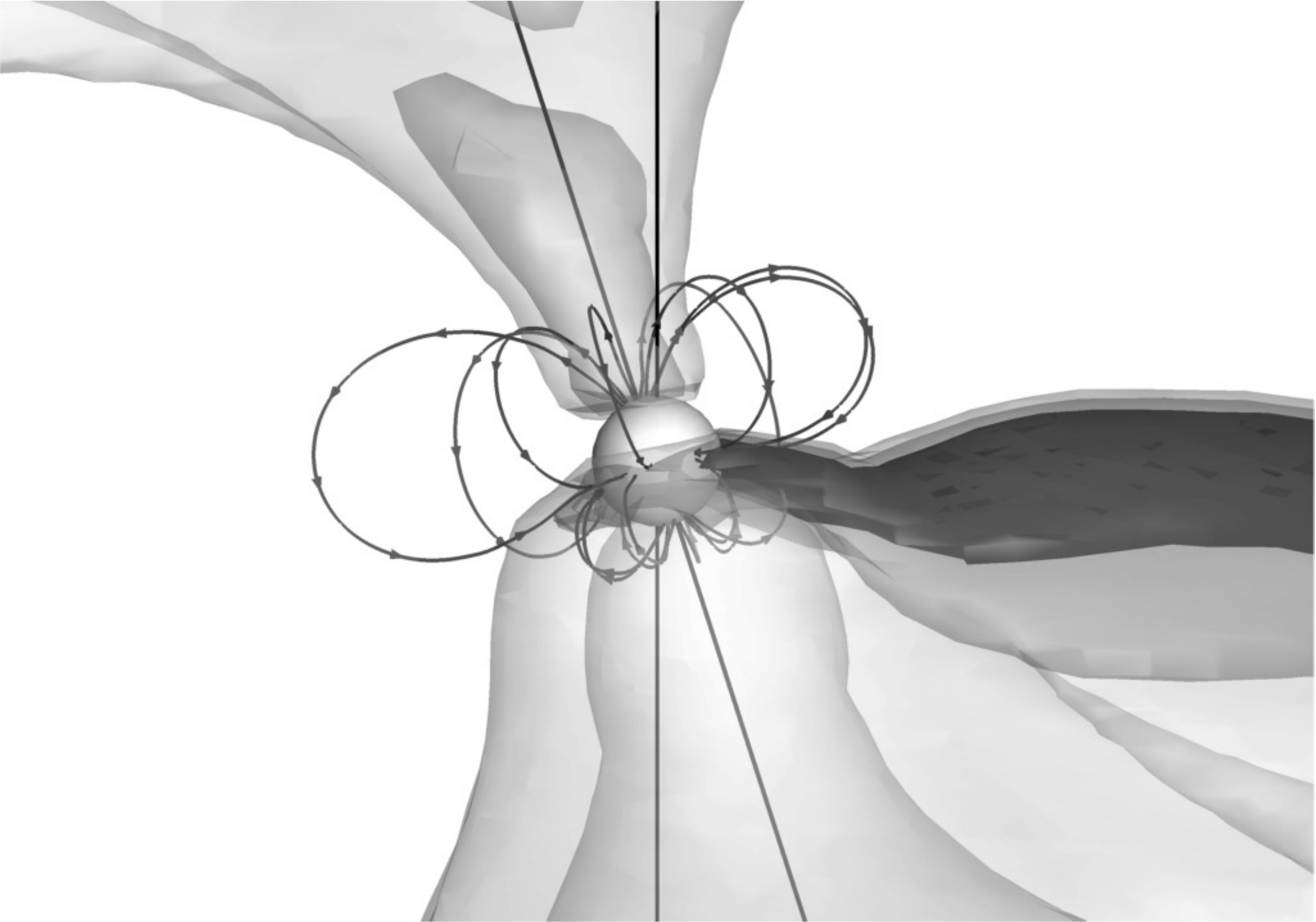}
\caption{3D structure for the phase $0$. The isosurfaces of the common logarithm of the 
density (in units of $\rho(L_1)$) for values $-6$, $-5$ and $-4$ and the 
magnetic field lines are shown. The gray scale of the field lines corresponds 
to the value of the magnetic induction. The straight line denotes the axis of 
rotation of the accretor. The inclined line is the magnetic axis. In the right 
panel the flow structure in the vicinity of the accretor is shown.}
\label{fg-3d-0}
\end{figure}

\begin{figure}[t]
\includegraphics[width=0.45\textwidth]{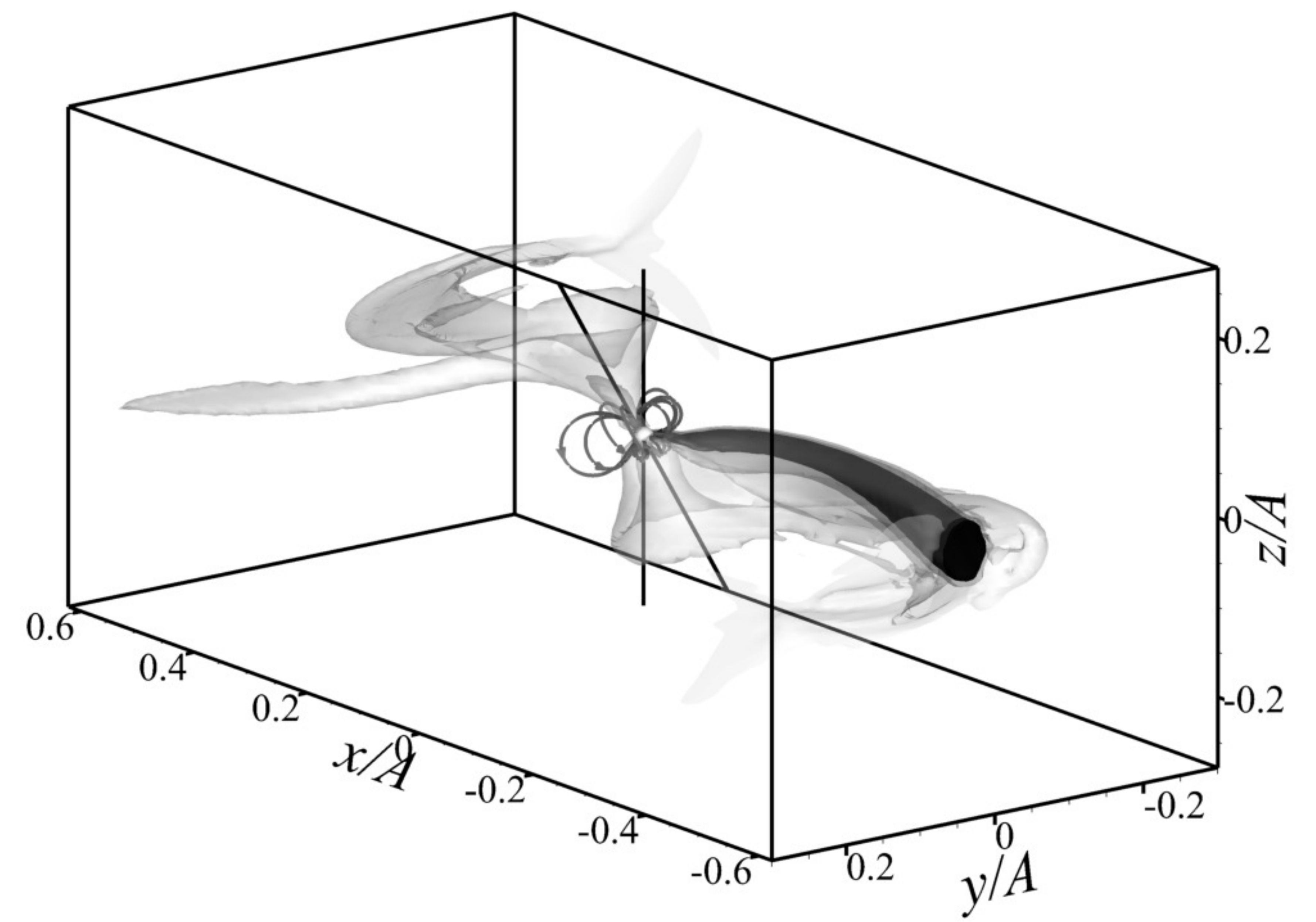}%
\hspace{0.1cm}
\includegraphics[width=0.45\textwidth]{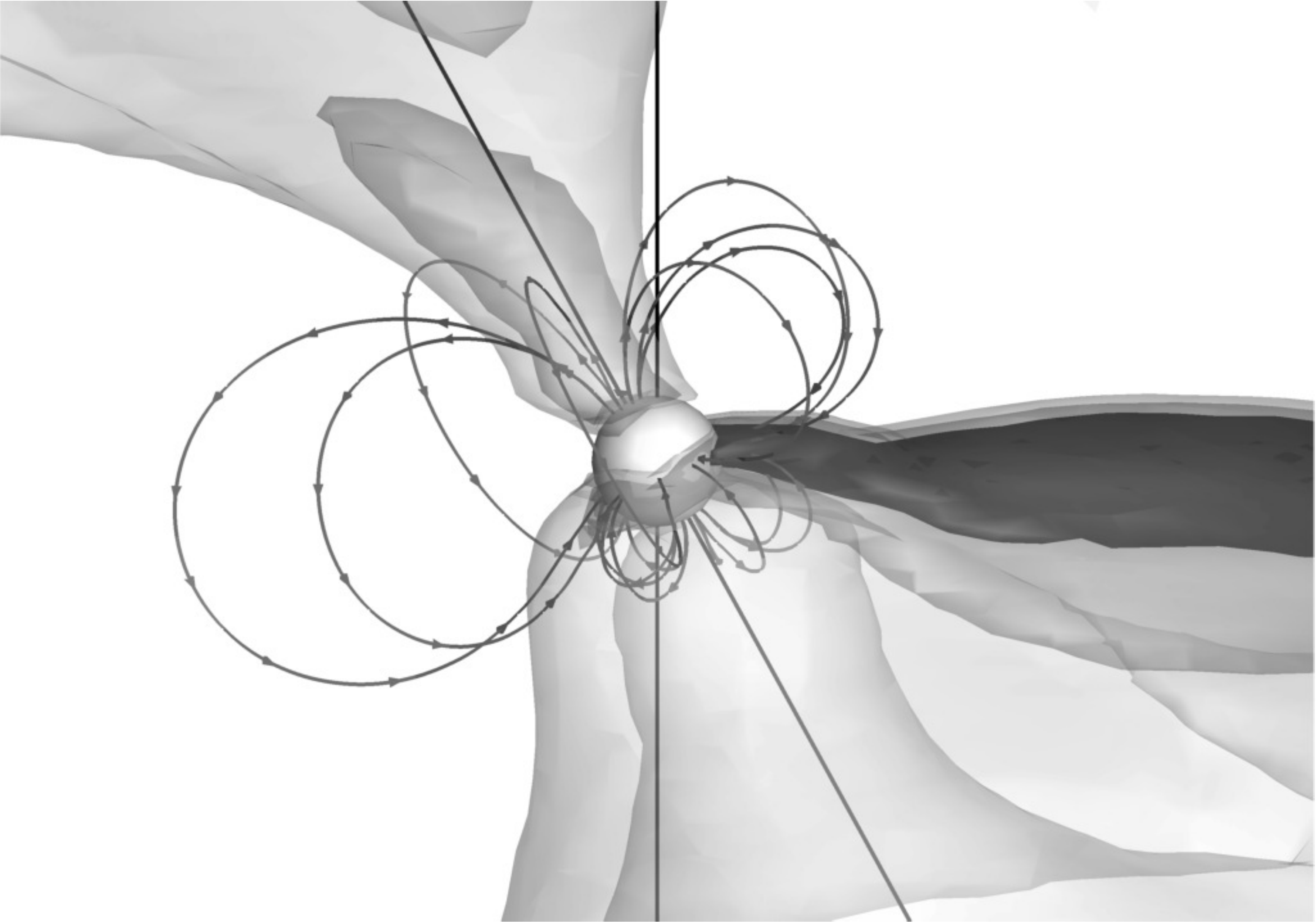}
\caption{The same as in Fig. \ref{fg-3d-0} but for the phase $0.1$.}
\label{fg-3d-1}
\end{figure}

\begin{figure}[b]
\includegraphics[width=0.45\textwidth]{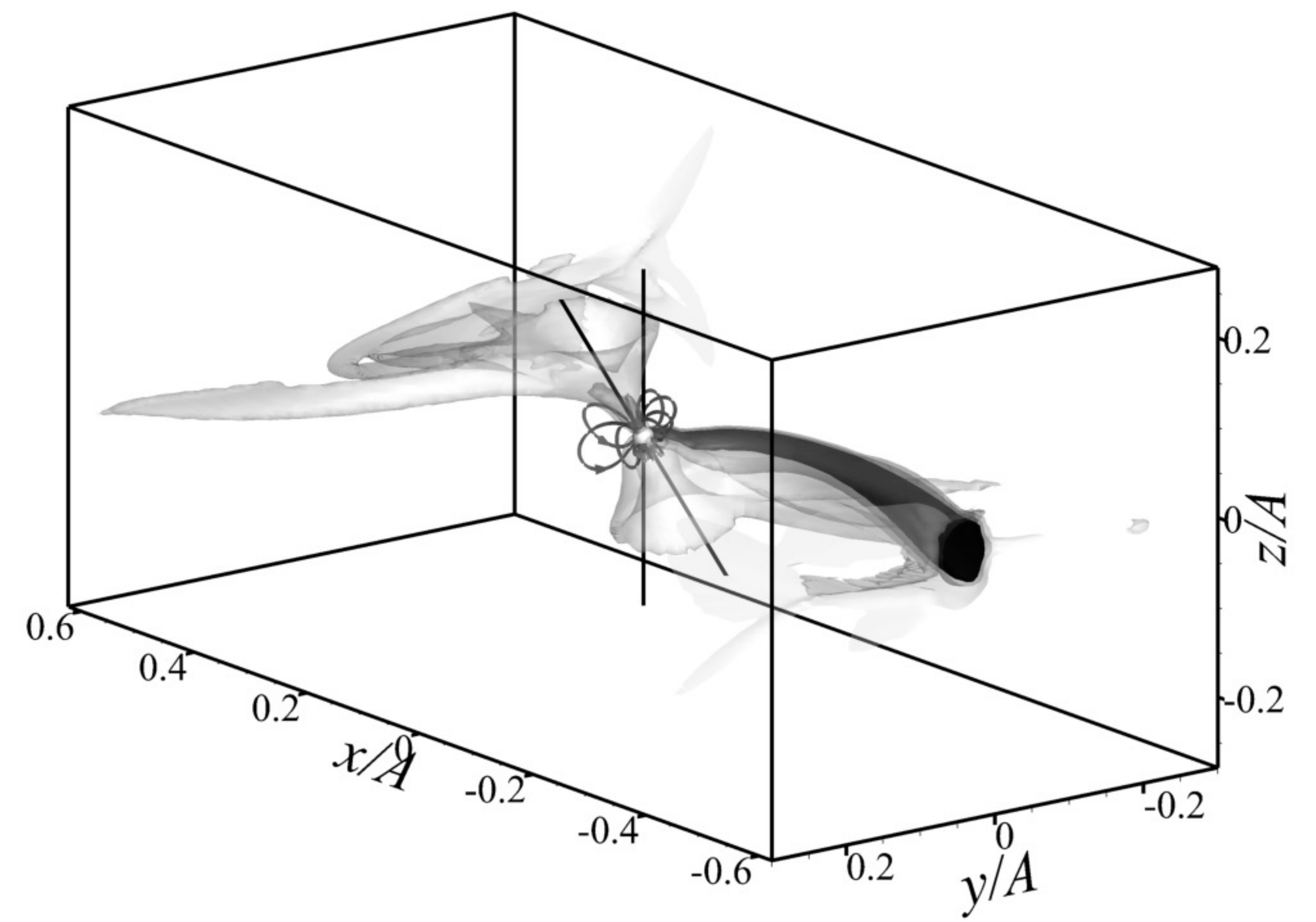}%
\hspace{0.1cm}
\includegraphics[width=0.45\textwidth]{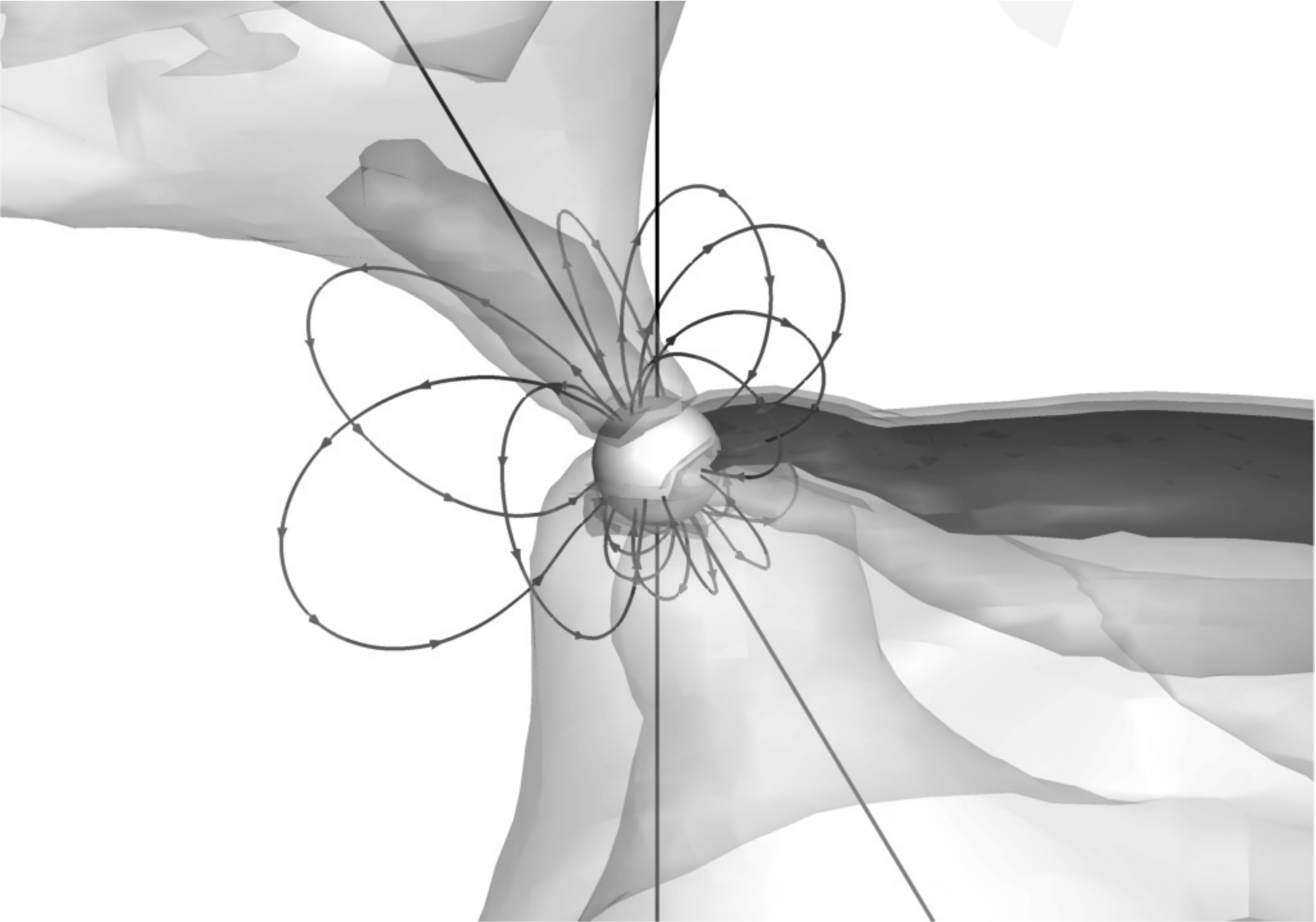}
\caption{The same as in Fig. \ref{fg-3d-0} but for the phase $0.2$.}
\label{fg-3d-2}
\end{figure}

\begin{figure}[t]
\includegraphics[width=0.45\textwidth]{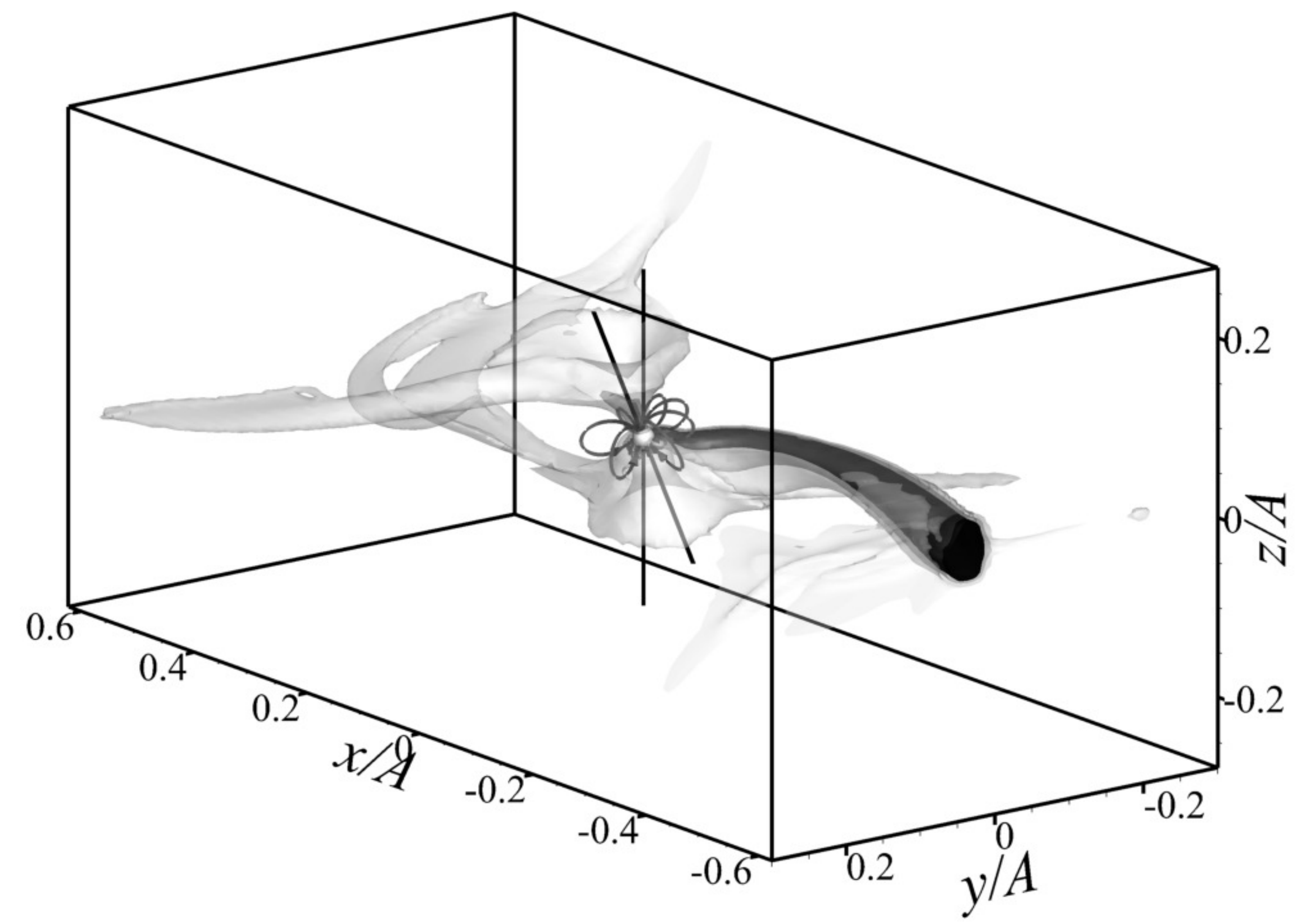}%
\hspace{0.1cm}
\includegraphics[width=0.45\textwidth]{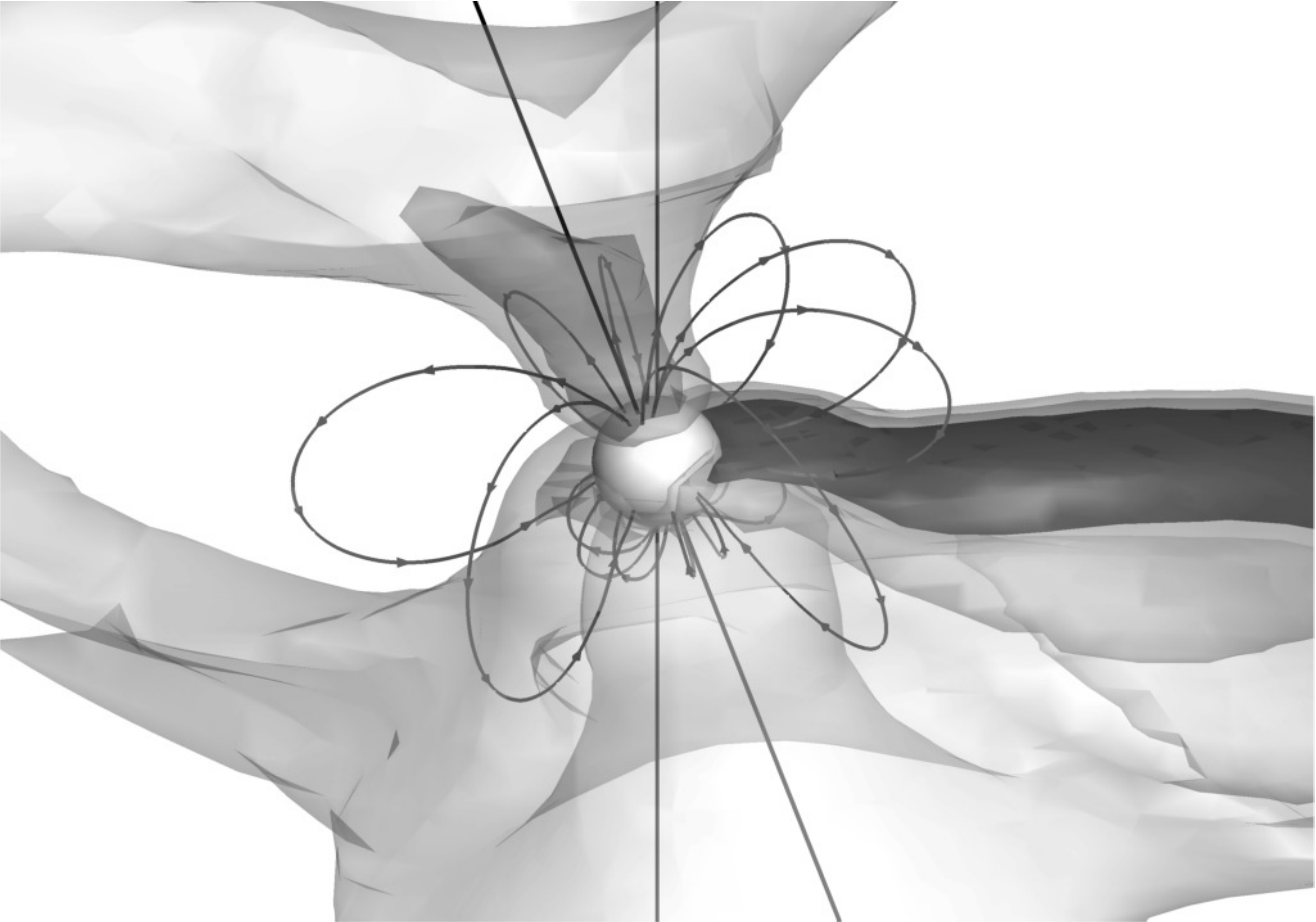}
\caption{The same as in Fig. \ref{fg-3d-0} but for the phase $0.3$.}
\label{fg-3d-3}
\end{figure}

\begin{figure}[b]
\includegraphics[width=0.45\textwidth]{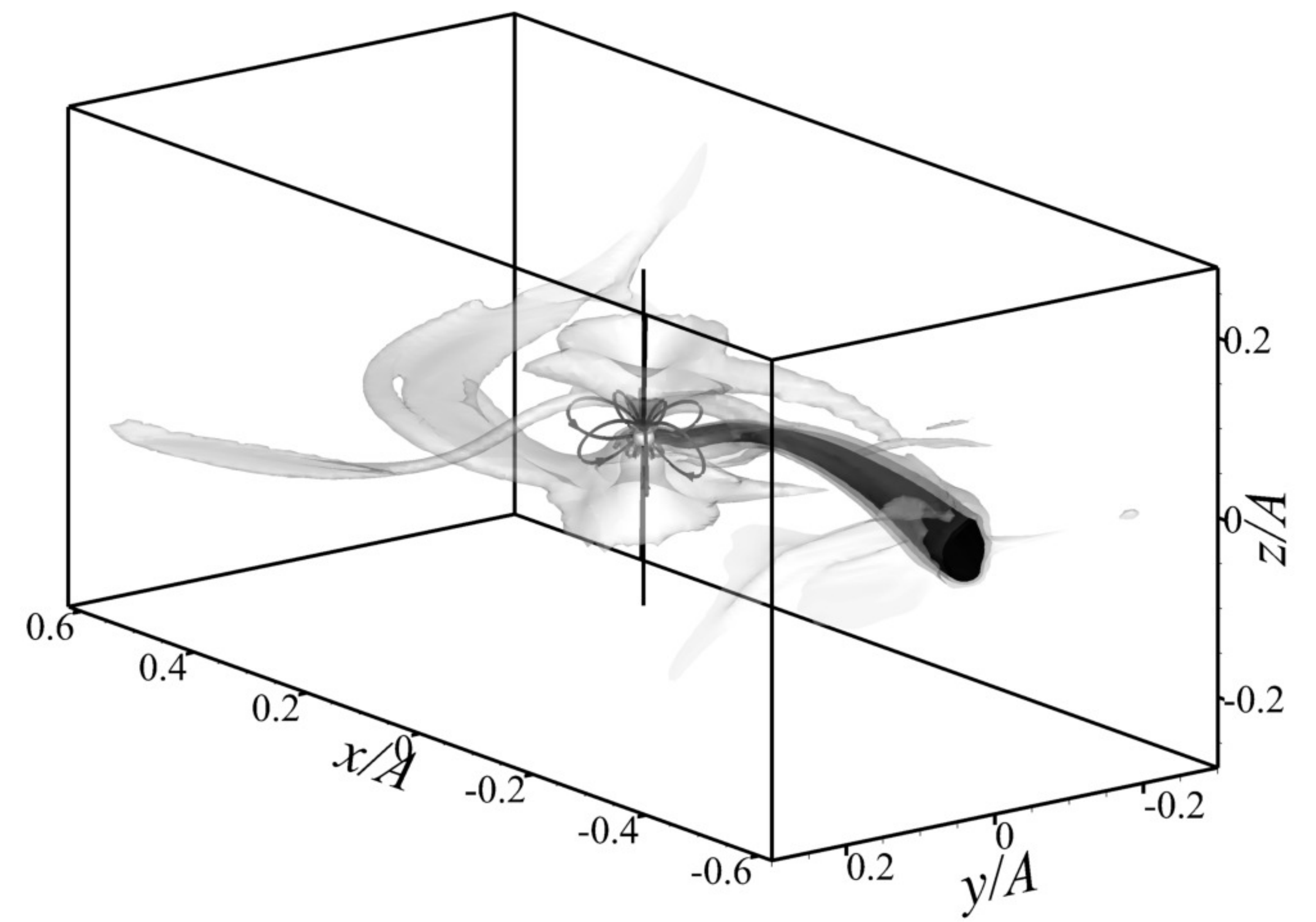}%
\hspace{0.1cm}
\includegraphics[width=0.45\textwidth]{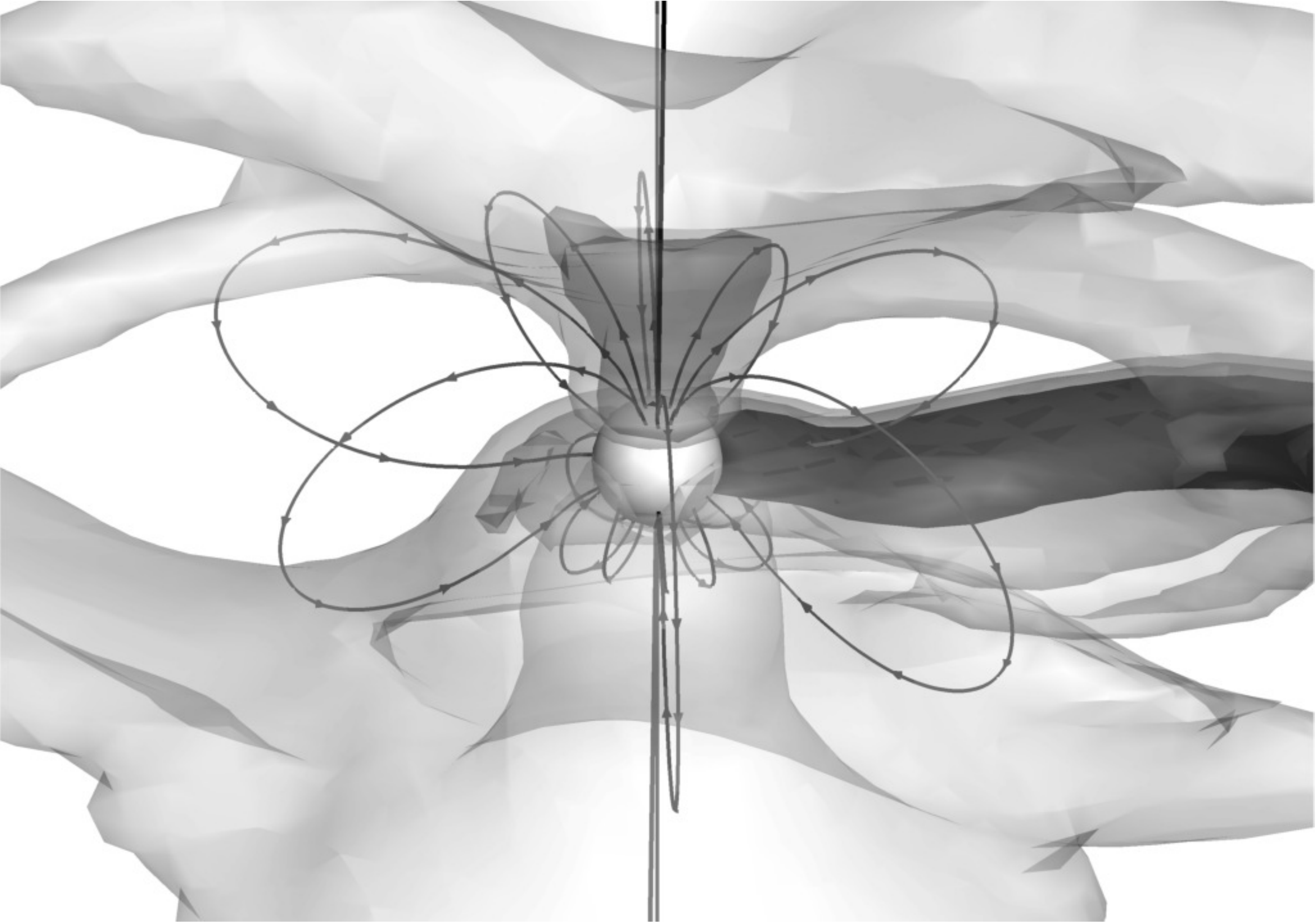}
\caption{The same as in Fig. \ref{fg-3d-0} but for the phase $0.4$.}
\label{fg-3d-4}
\end{figure}

\begin{figure}[t]
\includegraphics[width=0.45\textwidth]{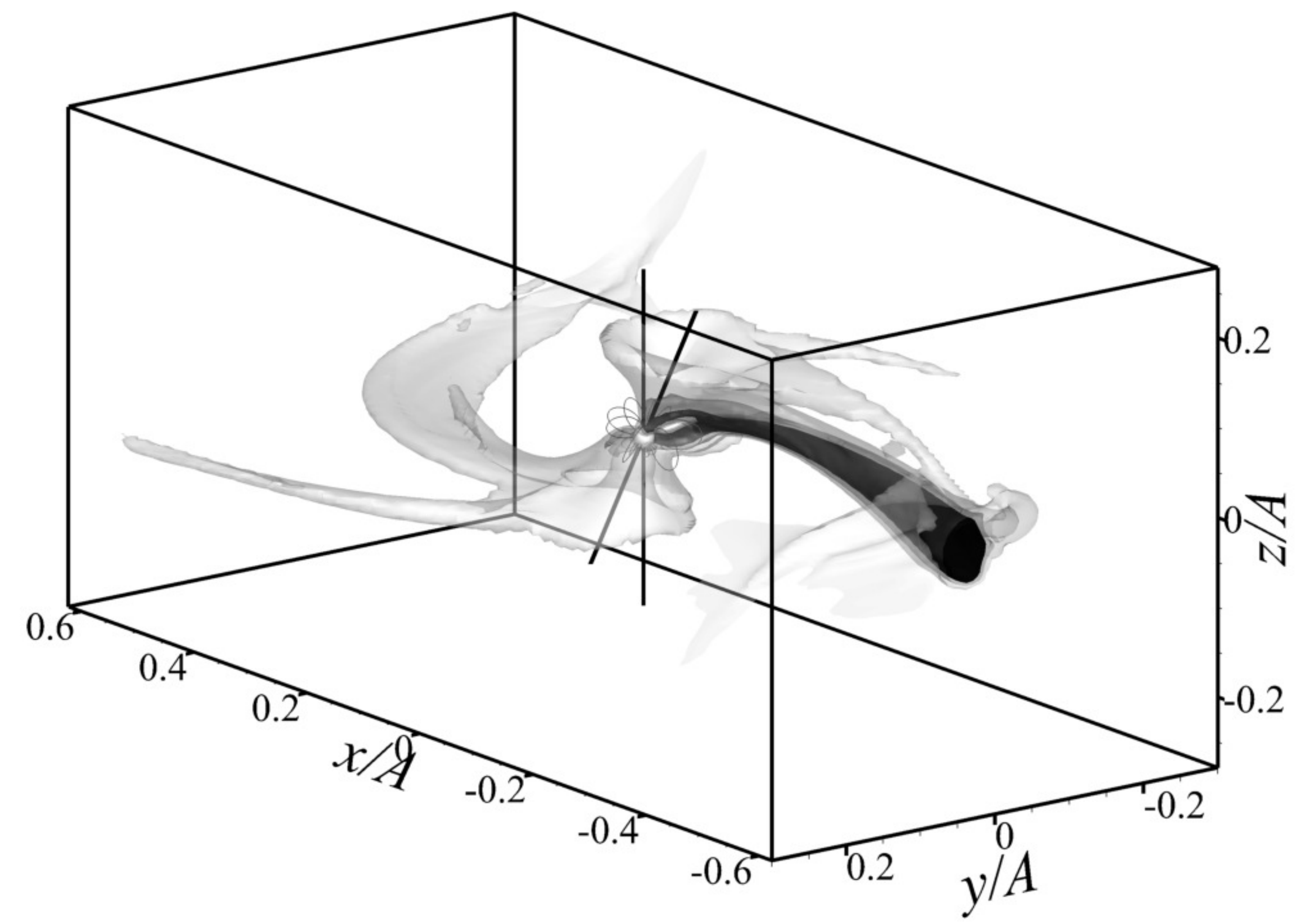}%
\hspace{0.1cm}
\includegraphics[width=0.45\textwidth]{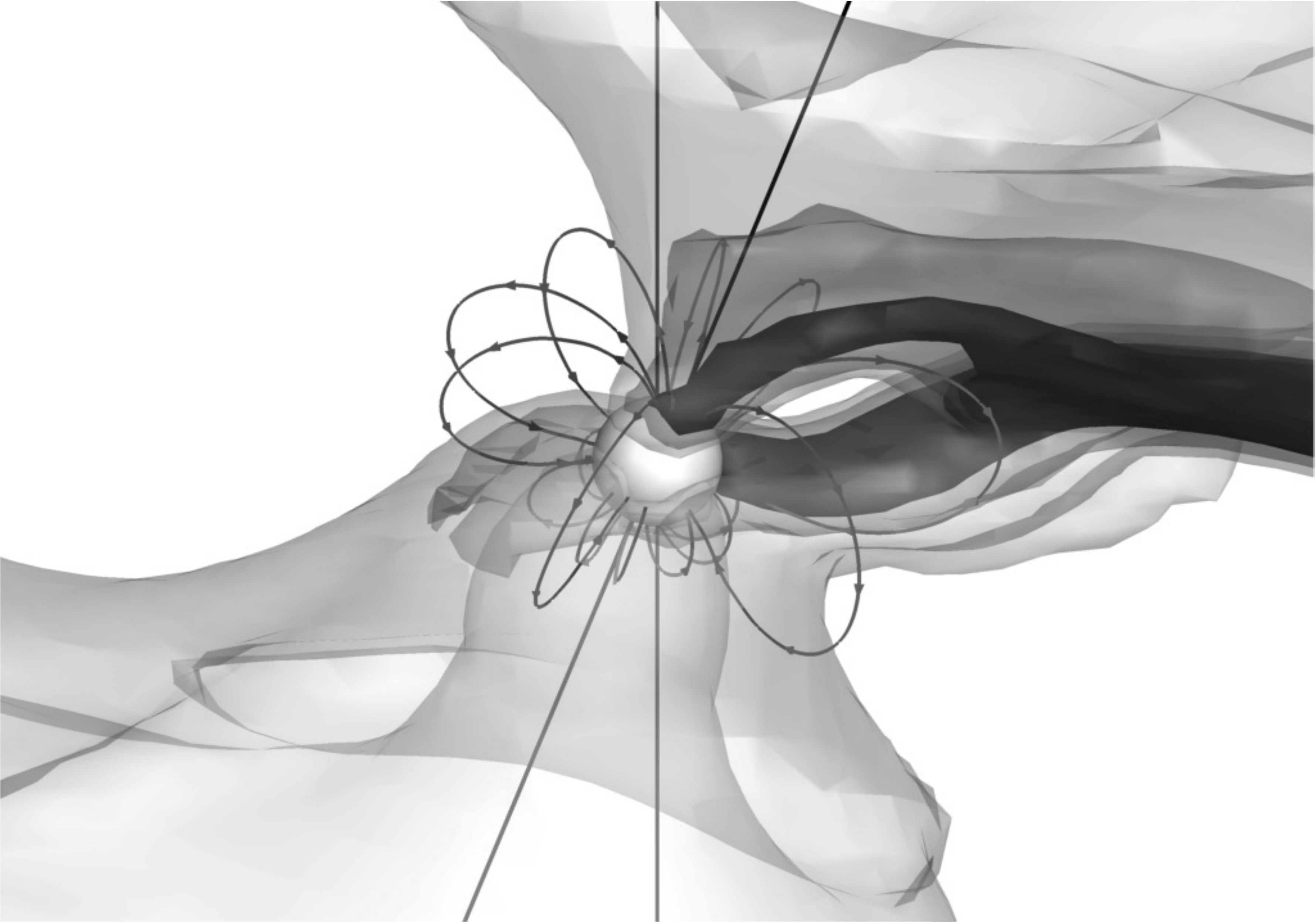}
\caption{The same as in Fig. \ref{fg-3d-0} but for the phase $0.5$.}
\label{fg-3d-5}
\end{figure}

\begin{figure}[b]
\includegraphics[width=0.45\textwidth]{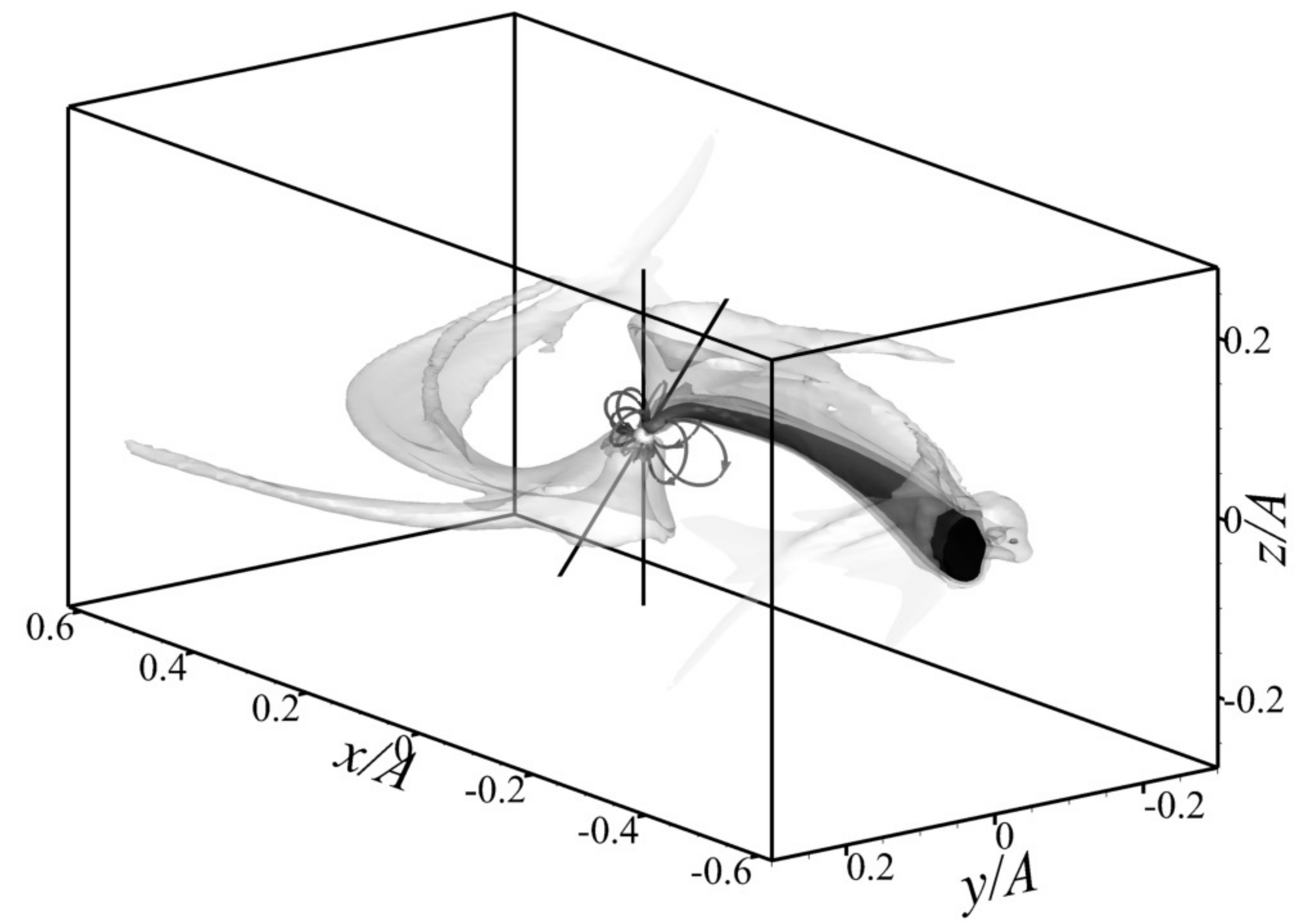}%
\hspace{0.1cm}
\includegraphics[width=0.45\textwidth]{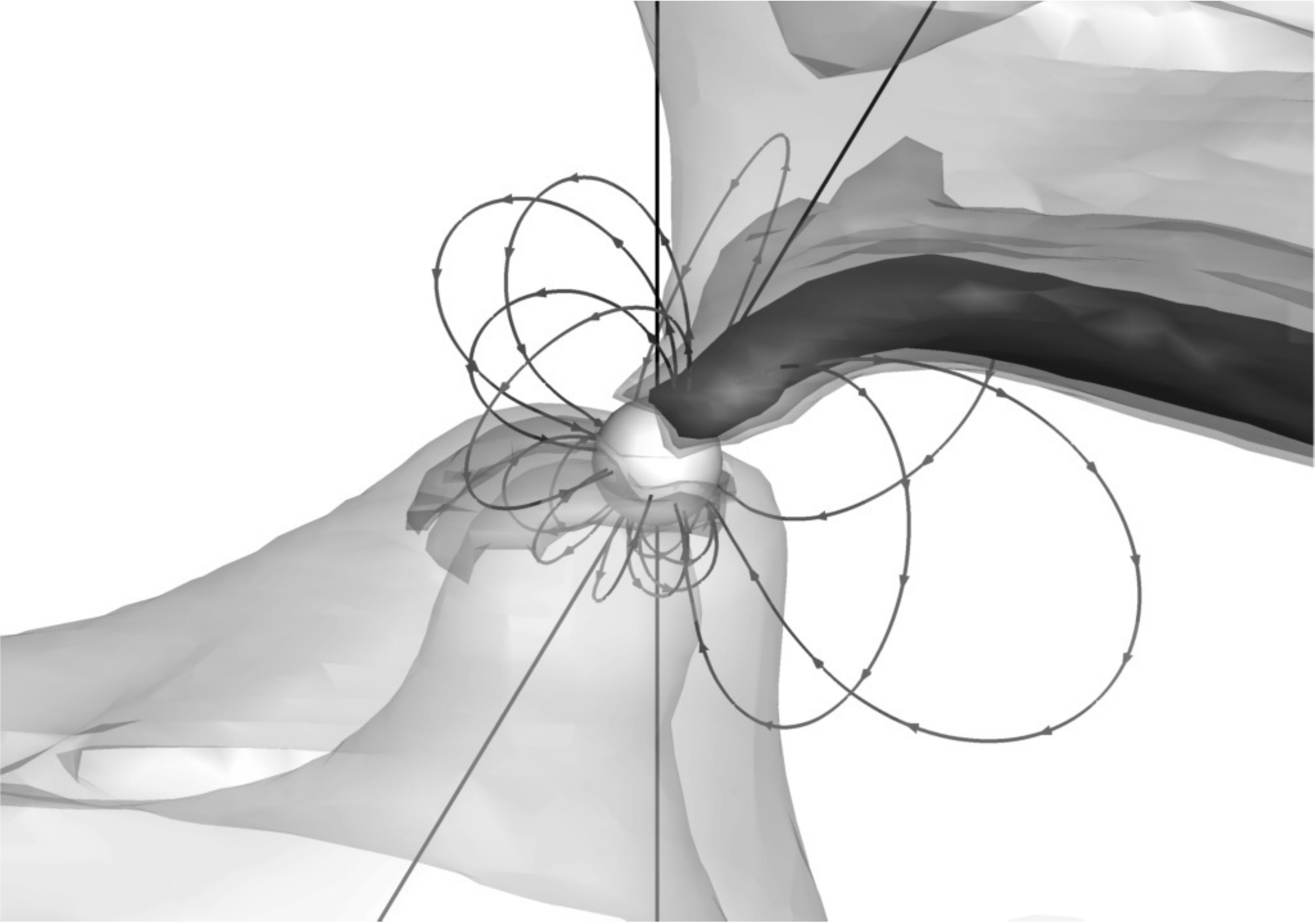}
\caption{The same as in Fig. \ref{fg-3d-0} but for the phase $0.6$.}
\label{fg-3d-6}
\end{figure}

\begin{figure}[t]
\includegraphics[width=0.45\textwidth]{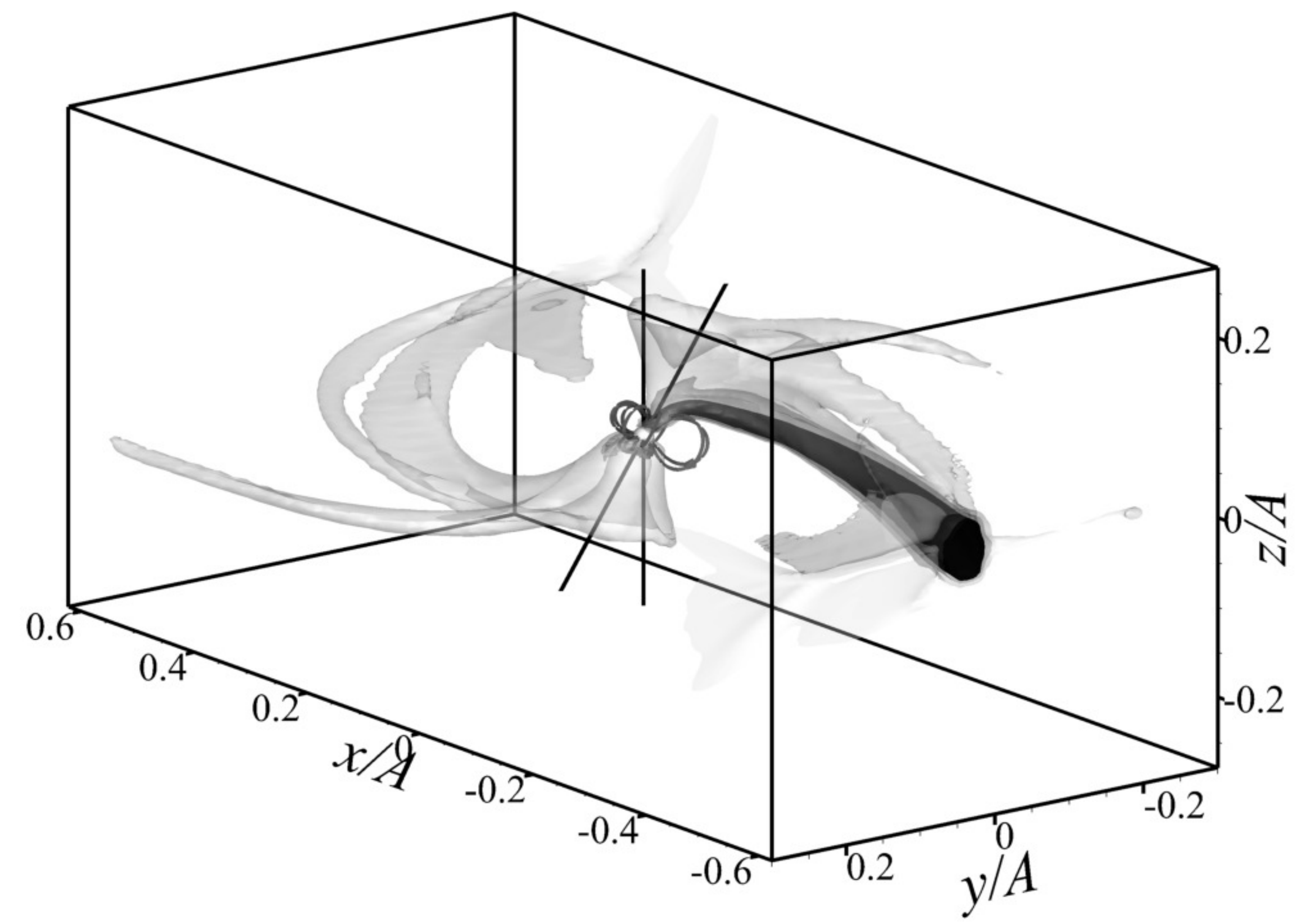}%
\hspace{0.1cm}
\includegraphics[width=0.45\textwidth]{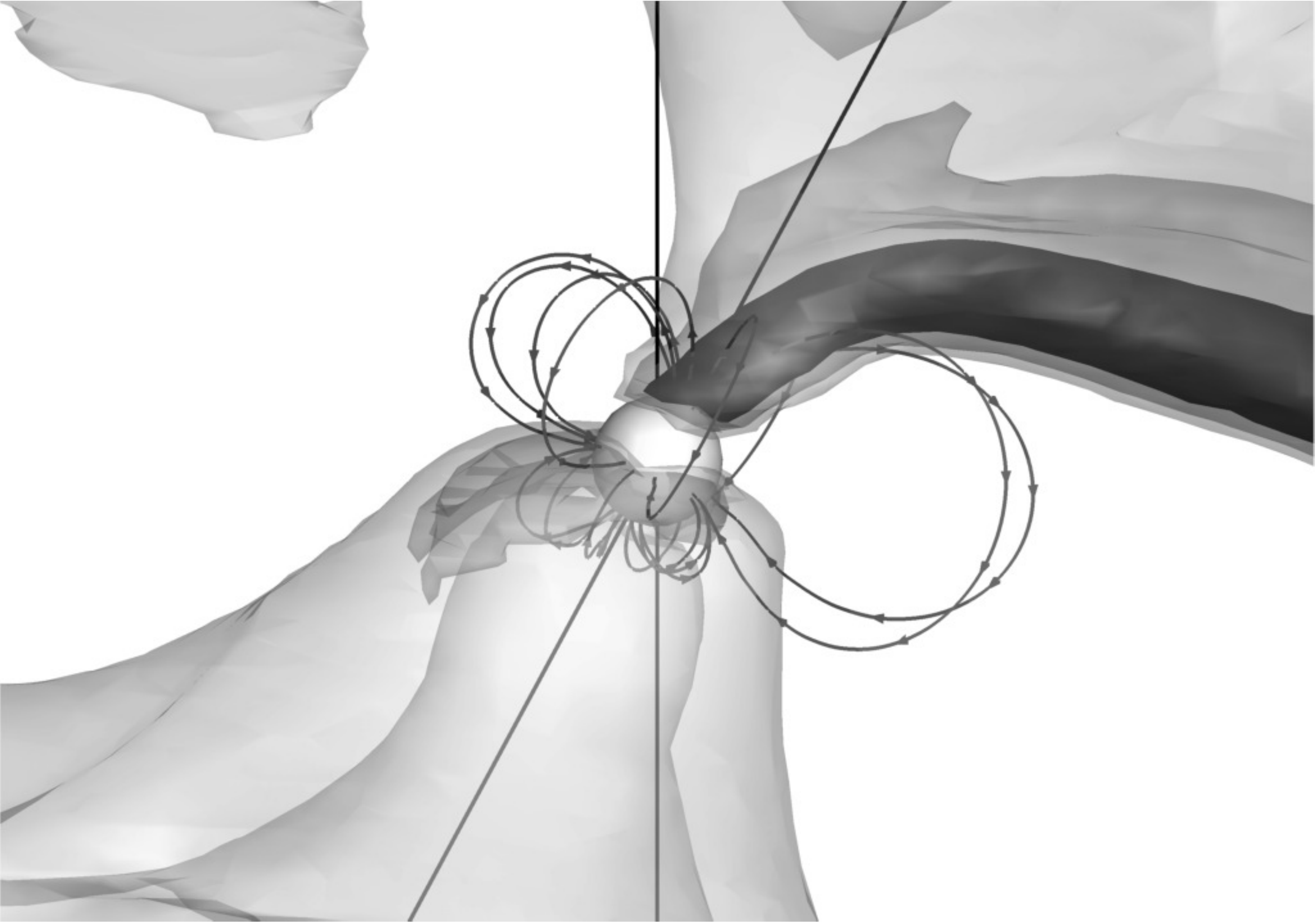}
\caption{The same as in Fig. \ref{fg-3d-0} but for the phase $0.7$.}
\label{fg-3d-7}
\end{figure}

\begin{figure}[b]
\includegraphics[width=0.45\textwidth]{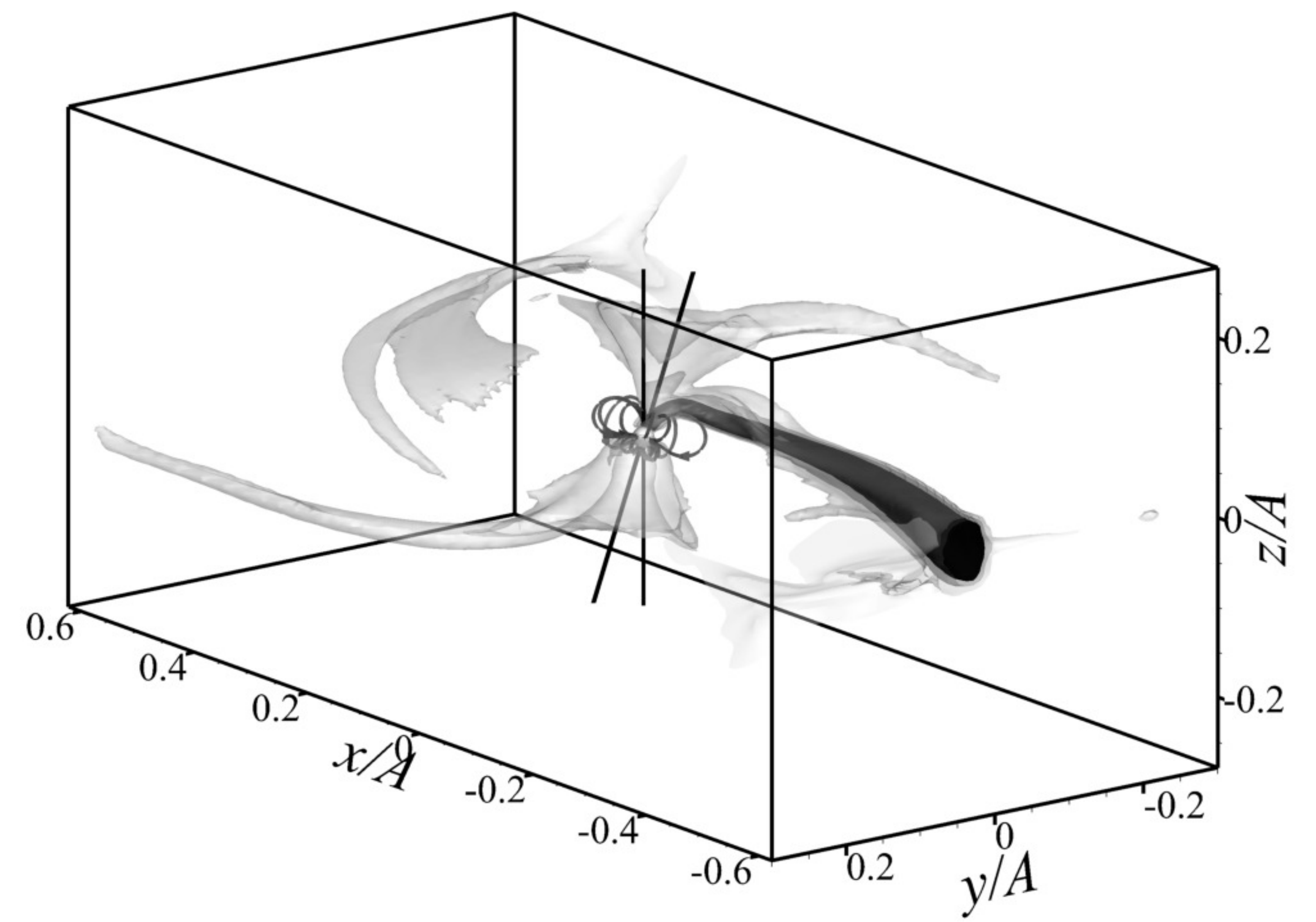}%
\hspace{0.1cm}
\includegraphics[width=0.45\textwidth]{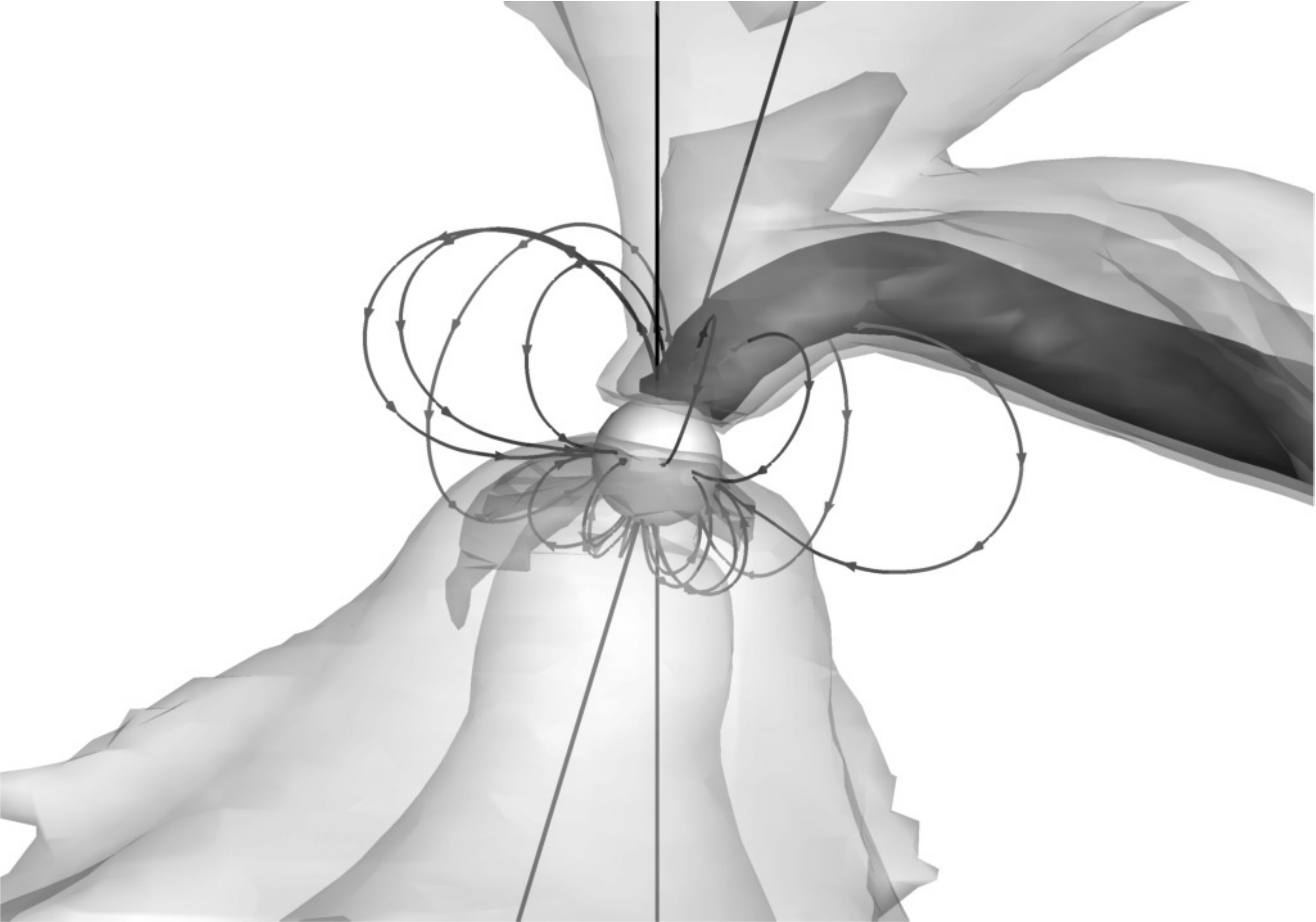}
\caption{The same as in Fig. \ref{fg-3d-0} but for the phase $0.8$.}
\label{fg-3d-8}
\end{figure}

\begin{figure}[t]
\includegraphics[width=0.45\textwidth]{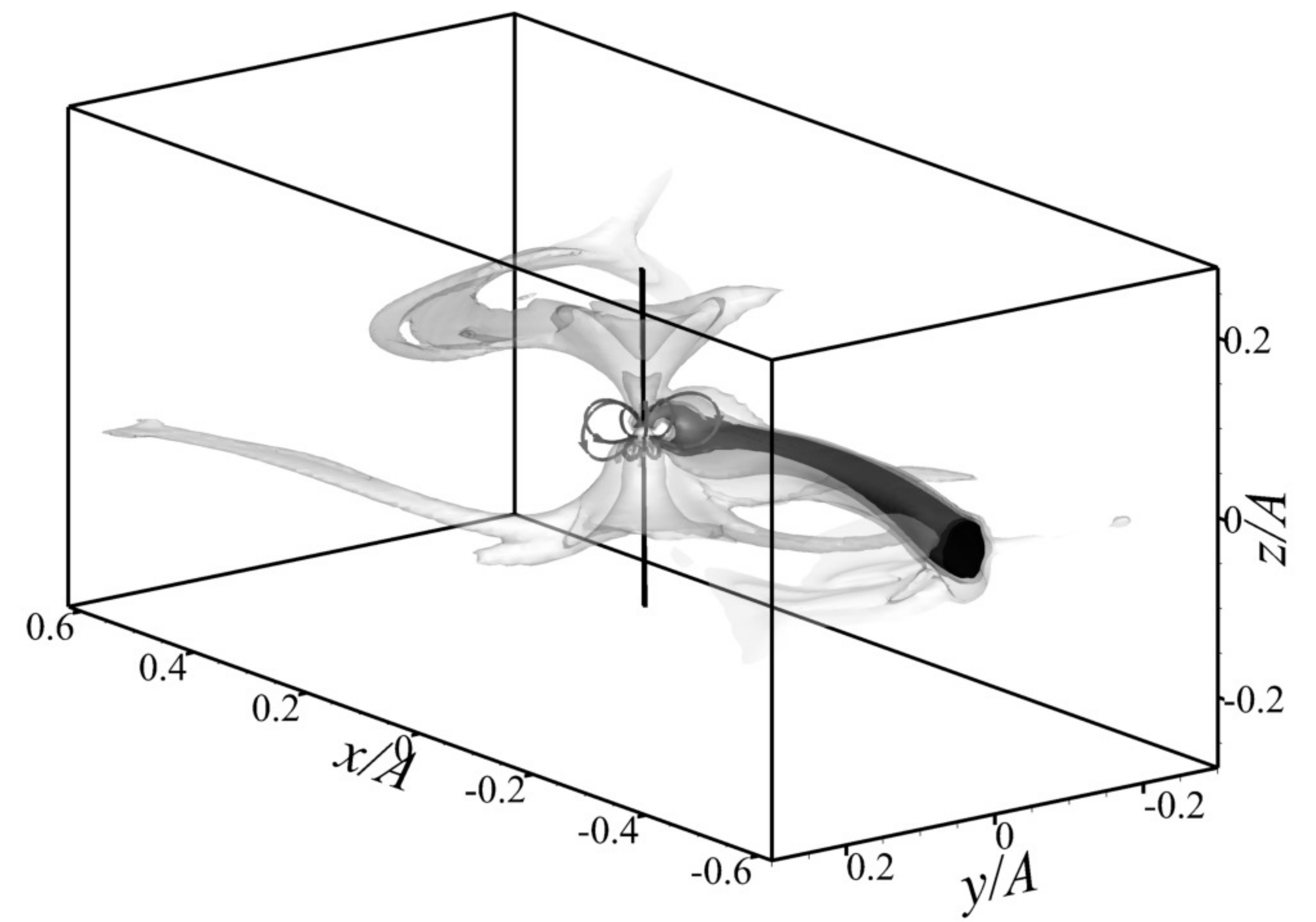}%
\hspace{0.1cm}
\includegraphics[width=0.45\textwidth]{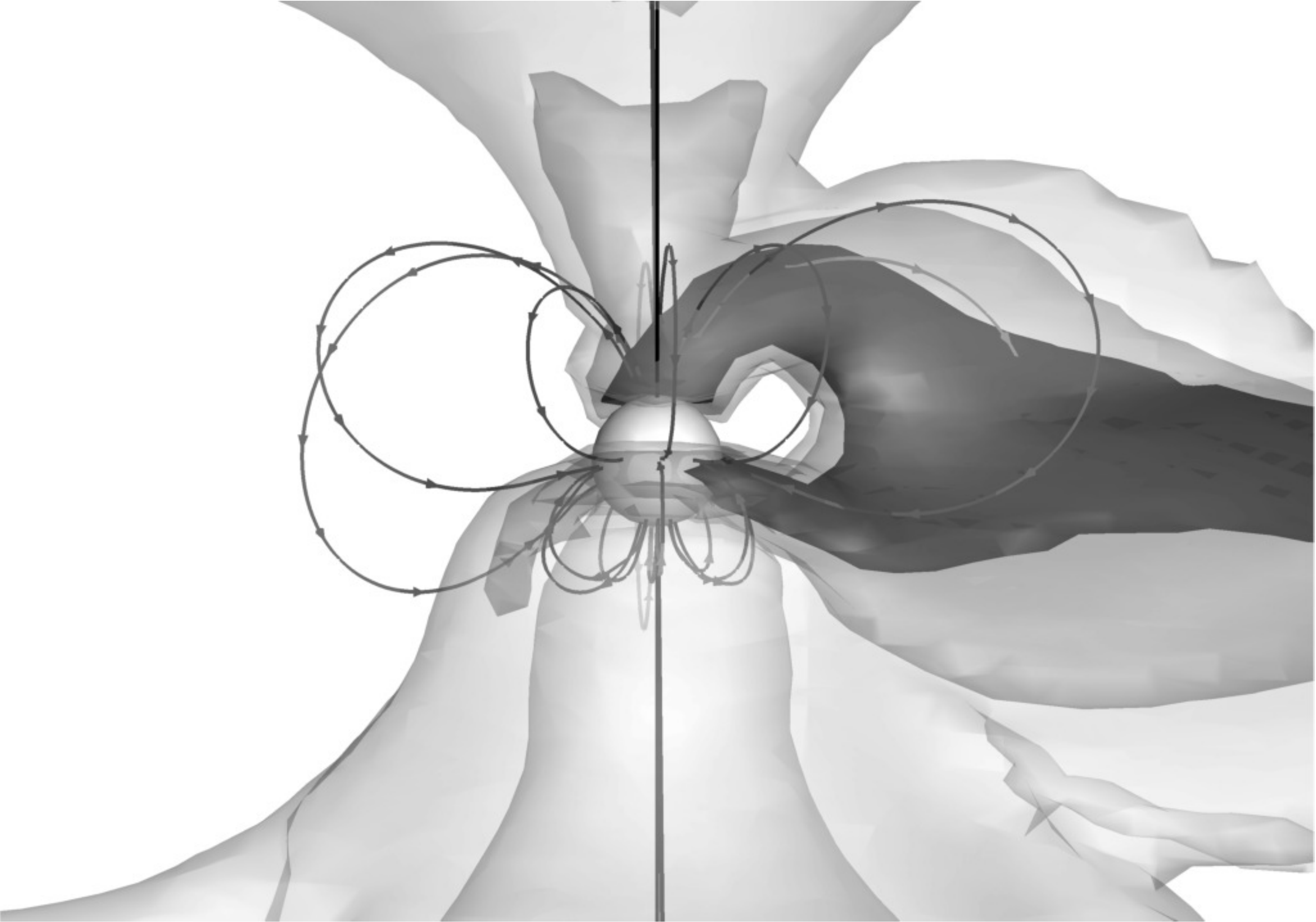}
\caption{The same as in Fig. \ref{fg-3d-0} but for the phase $0.9$.}
\label{fg-3d-9}
\end{figure}

The 3D flow structure obtained as a result of the calculations is shown in
Figures \ref{fg-3d-0}--\ref{fg-3d-9}. The isosurfaces of the density common
logarithm (in units of $\rho(L_1)$) for values $-6$, $-5$ and $-4$ are shown. 
The main stream is shown in a strongly contrasting manner. The magnetic field 
lines are denoted by the lines with the arrows. The gray scale of the lines 
corresponds to the induction of the magnetic field. The axis of the accretor's 
rotation (thin straight line) and magnetic axis (bold inclined line) are also 
shown. The left panels demonstrate the full flow pattern in the computational 
domain for a given polar or asynchronous polar beat phase. The right panels 
illustrate the flow structure in the vicinity of the accretor.

We should note that the structure of the accretion flows in the case of 
complex field geometry where the quadrupole component of the field exists 
significantly differs from the case of a purely dipole field. Calculations 
performed earlier \cite{ZBSMF2010} showed that in close binary systems where 
the accretor has a strong dipole magnetic field the accretion of matter may go 
either onto the northern magnetic pole or the southern pole or onto both of 
them depending on the value of the magnetic induction and orientation of the 
magnetic axis. Those calculations correspond to the standard one-pole vs. 
two-pole accretion geometries typically applied in models involving pure 
dipole fields. We show from the present calculations, that other modes of 
accretion may exist and even be common, if a complex magnetic field is 
present. 

Figures show that during the $P_\text{beat}$ period the flow structure in the
system significantly changes. At phase $0$ (Fig. \ref{fg-3d-0}) the accretion 
stream emanating from the inner Lagrangian point $L_1$ impacts the accretor in 
the magnetic belt whose existence is the result of the quadrupole component of 
the field. This leads to formation of an accretion ring surrounding the 
accretor. This is in a good agreement with the scheme demonstrated above (see 
Fig. \ref{fg-scheme}). In addition, the matter of the envelope forms an 
accretion stream falling onto the northern magnetic pole and a weaker stream 
flowing onto the southern magnetic pole. But these flows are much weaker than 
the main stream. Almost the same picture is seen at phases $0.1$ (Fig. 
\ref{fg-3d-1}), $0.2$ (Fig. \ref{fg-3d-2}) and $0.3$ (Fig.\ref{fg-3d-3}). The 
structure of the accretion flows at these phases differs only by the angle 
between the magnetic field axis and direction to the secondary component of the 
binary. These calculations demonstrate that the primary accretion zone in some 
polars may be near the spin equator, indicating the presence of a complex 
field.

At phase $0.4$ (Fig. \ref{fg-3d-4}) the flow structure changes. The main flow
of the accretion stream splits into two streams when approaching the accretor.
Most of the matter still falls onto the accretor's surface near the magnetic
belt. The second flow starts to move along the field lines toward the northern
magnetic pole. At phase $0.5$ (Fig. \ref{fg-3d-5}) this flow becomes much
stronger and intensities of the two flows become nearly equivalent. Thus, at
this moment two accretion zones of almost the same intensity form on the 
star's surface. One of them is located at the northern magnetic pole of the 
star, another one lies near the magnetic belt. Some polars may be accreting in 
this manner. For these binaries, the presence of a complex field will be 
apparent.

At phases $0.6$ (Fig. \ref{fg-3d-6}), $0.7$ (Fig. \ref{fg-3d-7}) and $0.8$
(Fig. \ref{fg-3d-8}) the accretion flow onto the magnetic belt runs out and
almost all the matter falls onto the northern magnetic pole. This flow
pattern is analogous to the case of polars where the accretor possesses a
purely dipole magnetic field \cite{ZBSMF2010}.

Finally, at phase $0.9$ (Fig. \ref{fg-3d-9}) the second change of the flow
structure during the $P_\text{beat}$ period occurs. The flow from the accretion 
stream again splits into two streams (as at the phase $0.5$). One flow still 
falls onto the northern magnetic pole and another forms an accretion region 
along the magnetic belt.

To summarize, we see that the accretion flow pattern is highly dependent on 
azimuthal angle. We find that among our ensemble of ten synchronized polars, 
single-pole accretion is most common. Often, the single-pole accretion mode 
occurred with flow onto a magnetic pole. This accretion mode will likely be 
difficult to observationally differentiate from standard dipole accretion. 
Another common mode found involves a single equatorial spot. The position of 
the equatorial spot varies somewhat as a function of azimuthal angle. In a 
minority of cases our model resulted in a two-pole accretion mode. In this 
mode, accretion takes place onto a magnetic pole and on a spot just south of 
the spin-equator simultaneously. The identification of such a two-pole mode 
would place strong constrains on magnetic field structures, such as the one 
modelled in this paper.

We move to the question of modelling BY Cam. It is clear that during the 
$P_\text{beat}$ period the flow structure changes twice. At those times, the 
accretion stream configuration and the number and location of accretion zones 
change. This is relatively consistent with the pole switching model of
\cite{Mason1995}. However, in that model pole switching takes place between 
diametrically opposed equatorial poles. It is not yet clear if the pole 
switching mode from a magnetic pole to an equatorial spot unveiled here may be 
applied to BY Cam. We expect that other system parameters, such as accretion 
rate, mass ratio, and magnetic field strength, not explored in the current 
paper will have significant affects on accretion geometry and may be modelled
in the manner described here. In fact, additional accretion modes may be 
possible.

\subsection{Hot spots}

To calculate locations of hot spots forming as zones of intense accretion we
used a method that was described by Romanova et al. \cite{Romanova2004}. It is 
assumed that when matter falls onto the accretor its thermal and kinetic 
energy turns into radiation. The density of the energy flux through the 
accretor's surface at a given point $\vec{r} = \vec{R}$ is determined by the 
relation:
\begin{equation}\label{eq13}
 f(\vec{R}) = -\rho \vec{n} \cdot \vec{v}
 \left( \varepsilon + {\vec{v}^2}/{2} + {P}/{\rho} \right), 
\end{equation}
where $\varepsilon$ is the internal energy of gas per unit mass and $\vec{n}$
is the unit vector normal to the surface. The sign ''$-$'' has been chosen
since on the accretor's surface the normal component of the velocity is
$\vec{n} \cdot \vec{v} < 0$. As a result the value $f(\vec{R})$ appears
positive in \eqref{eq13}.

For the sake of simplicity we assume that the radiation coming from the 
accretion zones is of the black body type. This means that the local effective 
temperature at a given point of the surface satisfies the relation: 
$f(\vec{R}) = \sigma T^4_\text{eff}(\vec{R})$, where $\sigma$ is the 
Stefan-Boltzmann constant. We should note that in our work matter is accreted 
directly from the accretion stream but not from the disk as in 
\cite{Romanova2004}. Thus the normal component of the velocity over the entire 
surface of the accretor with good precision is equal to the free fall 
velocity,
\begin{equation}\label{eq14}
 \vec{n} \cdot \vec{v} = -v_\text{ff} = 
 -\sqrt{{2G M_\text{a}}/{R_\text{a}}}.
\end{equation}
Therefore, the distribution of the energy flux density $f(\vec{R})$ and, 
hence, the effective temperature $T_\text{eff}(\vec{R})$ on the accretor's
surface will be determined by the density distribution $\rho(\vec{R})$.

\begin{figure}[p]
\centering
\begin{tabular}{c|c}
\hbox{\includegraphics[width=0.45\textwidth]{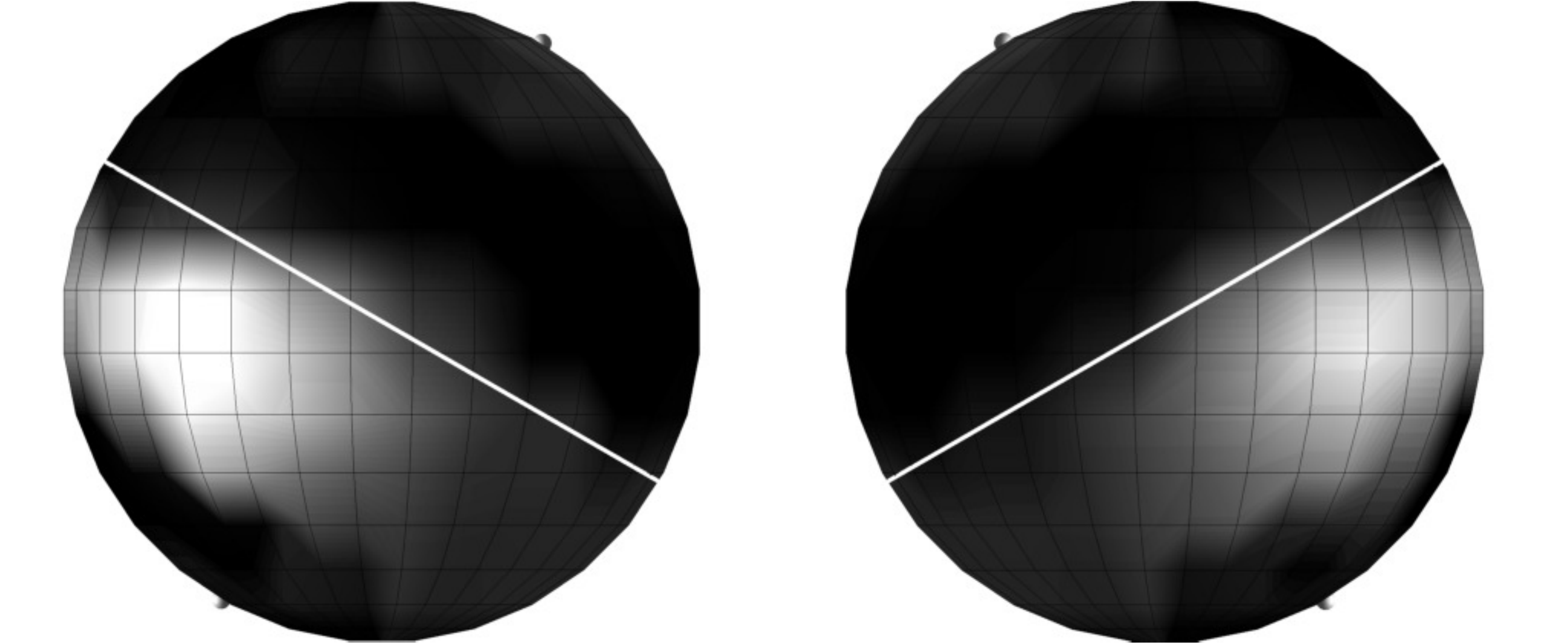}} & 
\hbox{\includegraphics[width=0.45\textwidth]{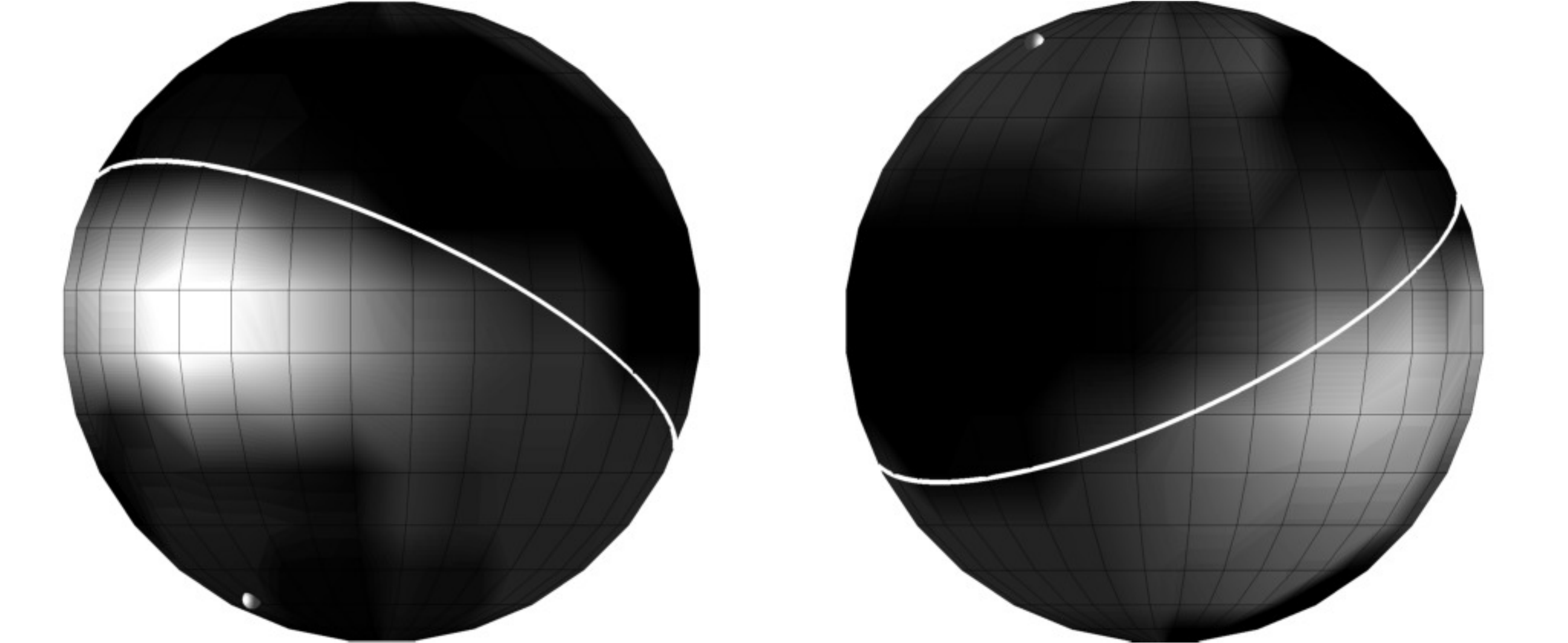}} \\
a) 0.0 & b) 0.1 \\ \hline
\hbox{\includegraphics[width=0.45\textwidth]{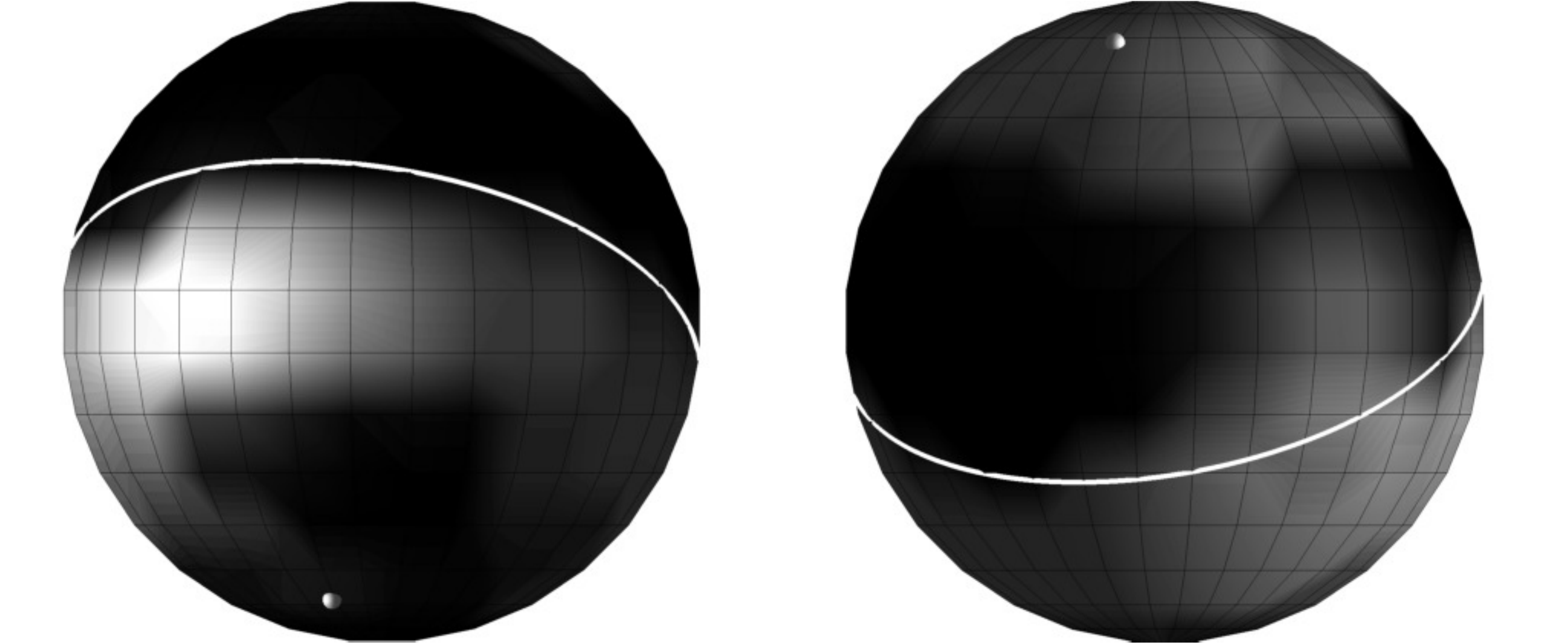}} & 
\hbox{\includegraphics[width=0.45\textwidth]{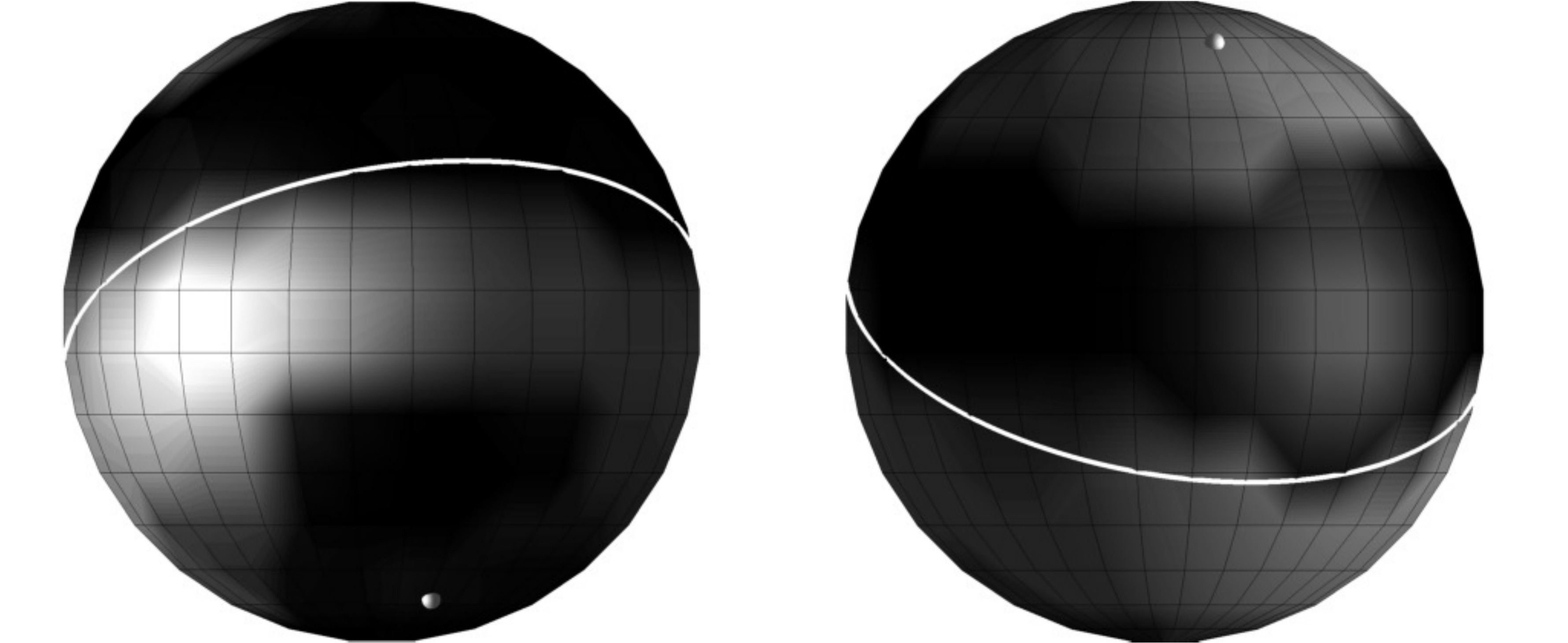}} \\
c) 0.2 & d) 0.3 \\ \hline
\hbox{\includegraphics[width=0.45\textwidth]{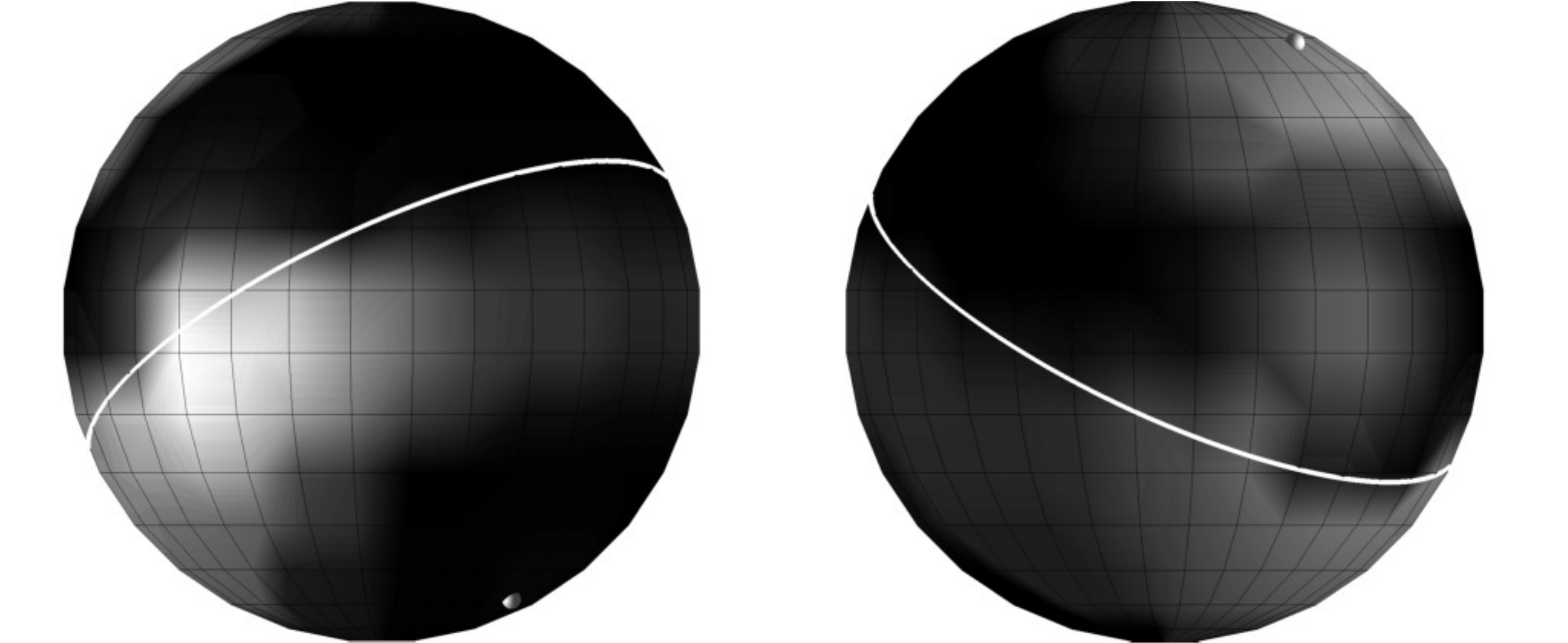}} & 
\hbox{\includegraphics[width=0.45\textwidth]{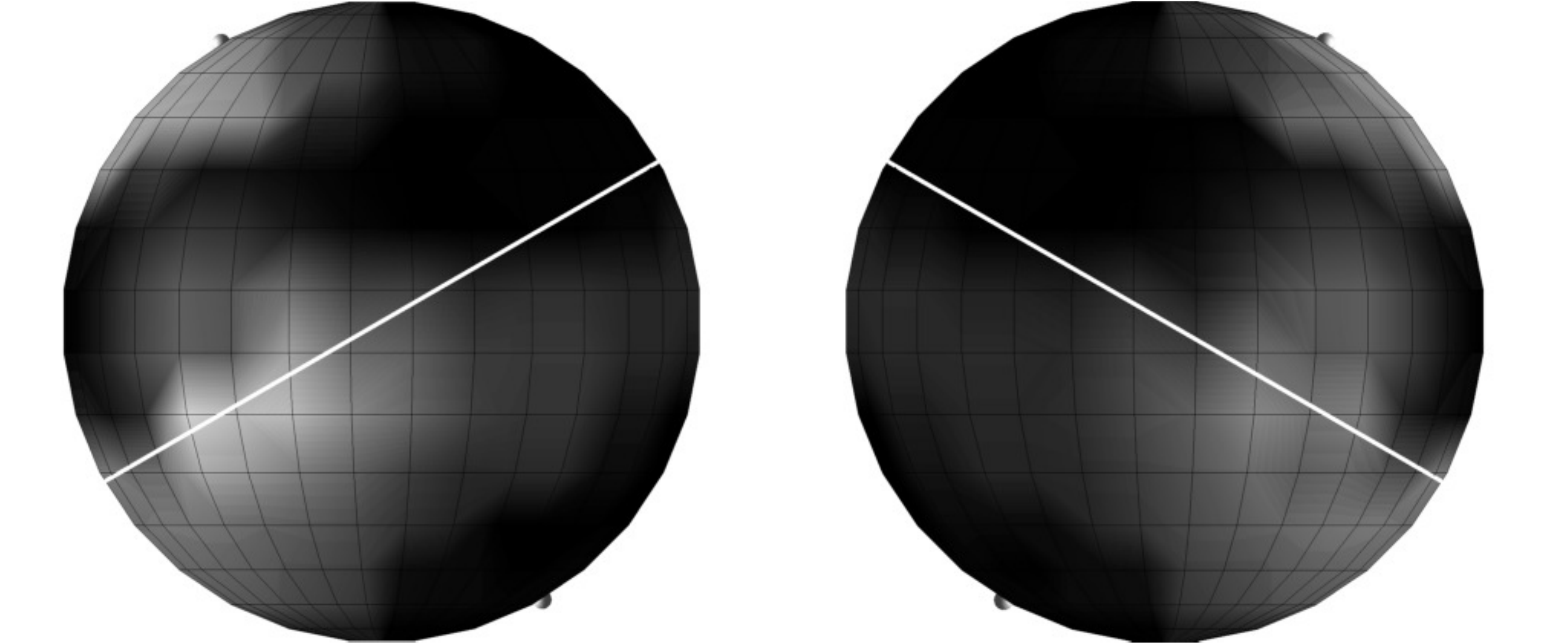}} \\
e) 0.4 & f) 0.5 \\ \hline
\hbox{\includegraphics[width=0.45\textwidth]{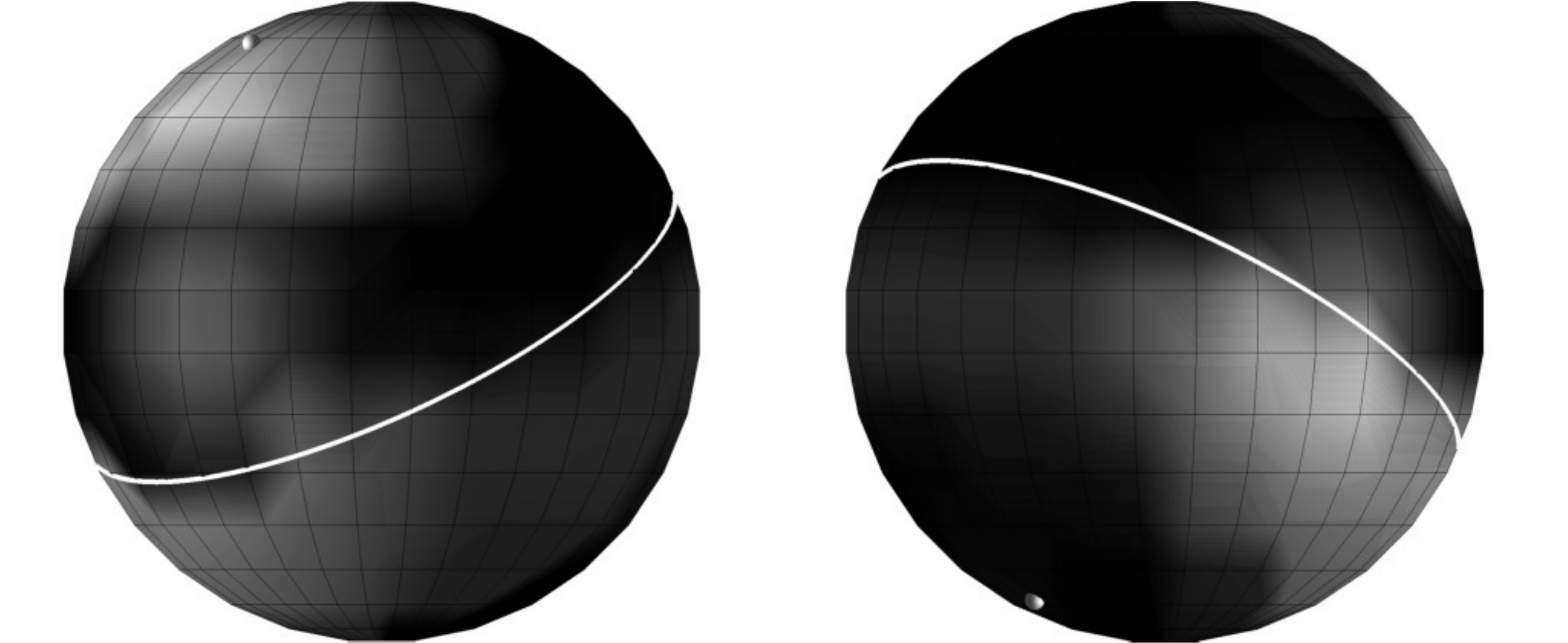}} & 
\hbox{\includegraphics[width=0.45\textwidth]{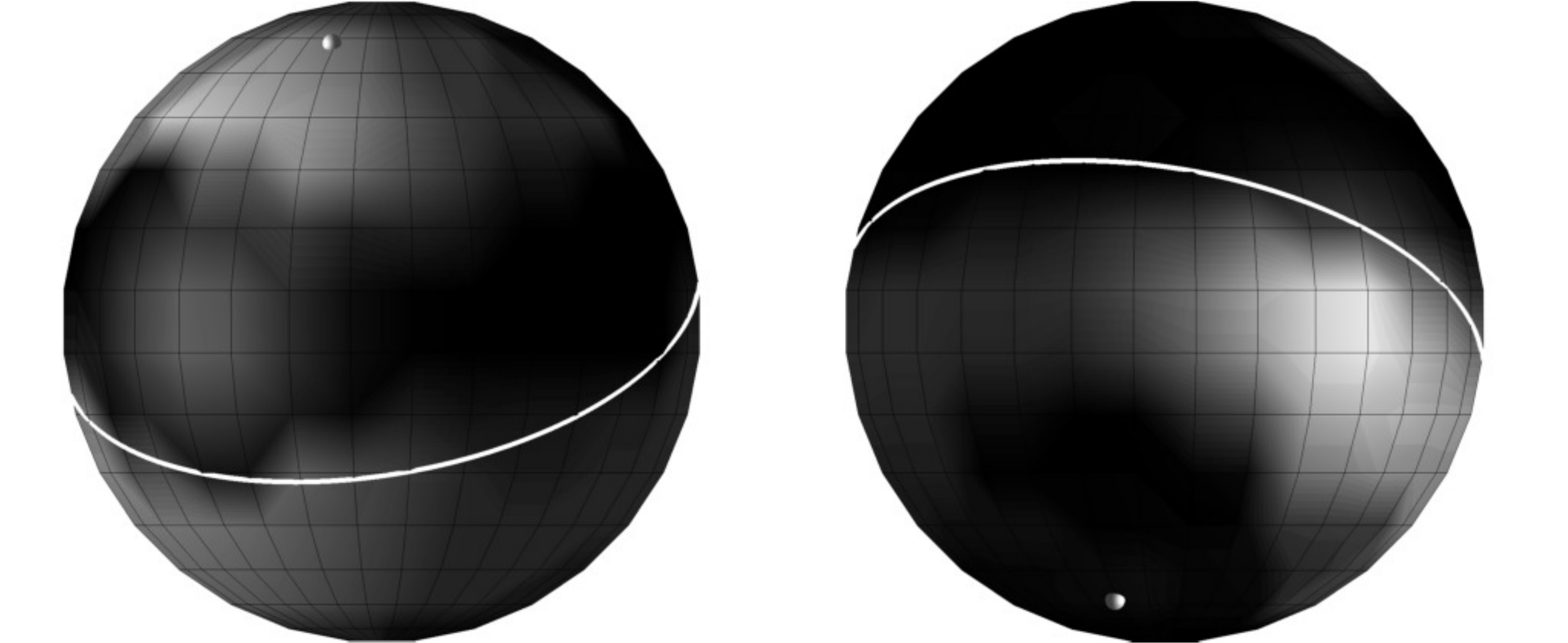}} \\
g) 0.6 & h) 0.7 \\ \hline
\hbox{\includegraphics[width=0.45\textwidth]{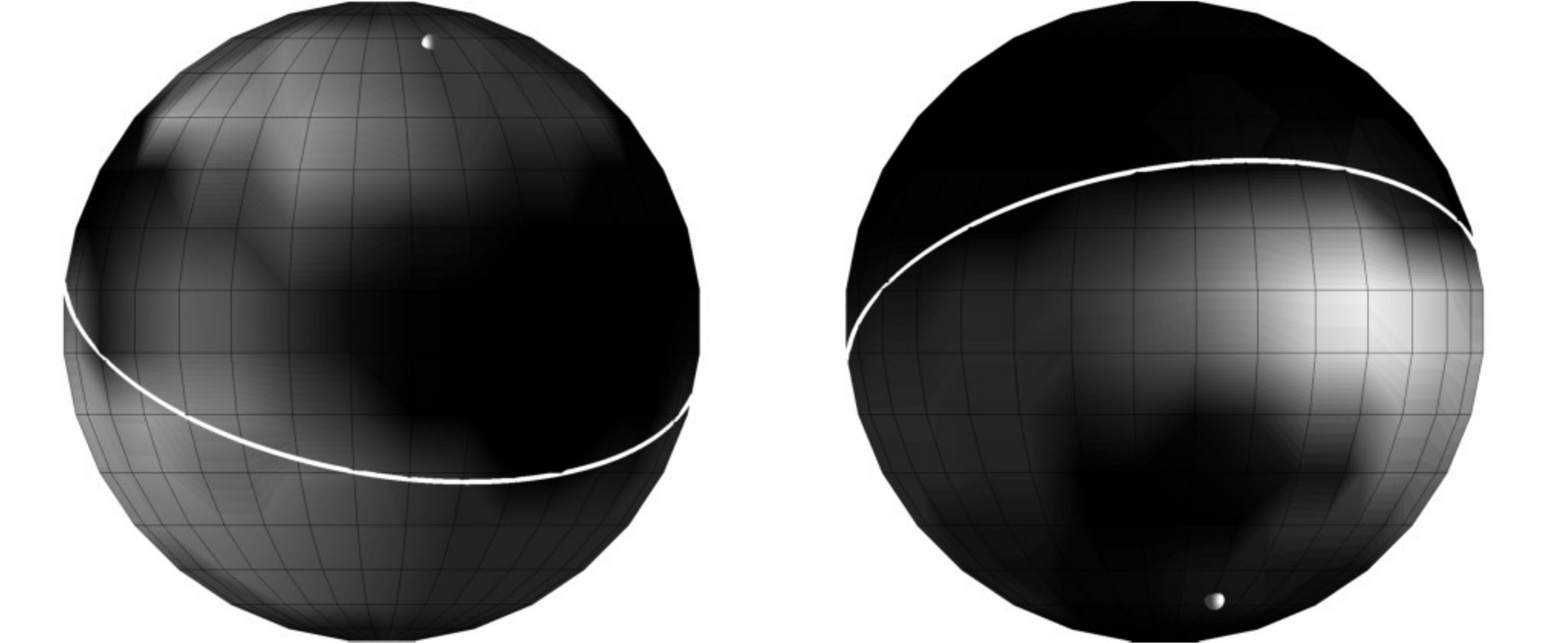}} & 
\hbox{\includegraphics[width=0.45\textwidth]{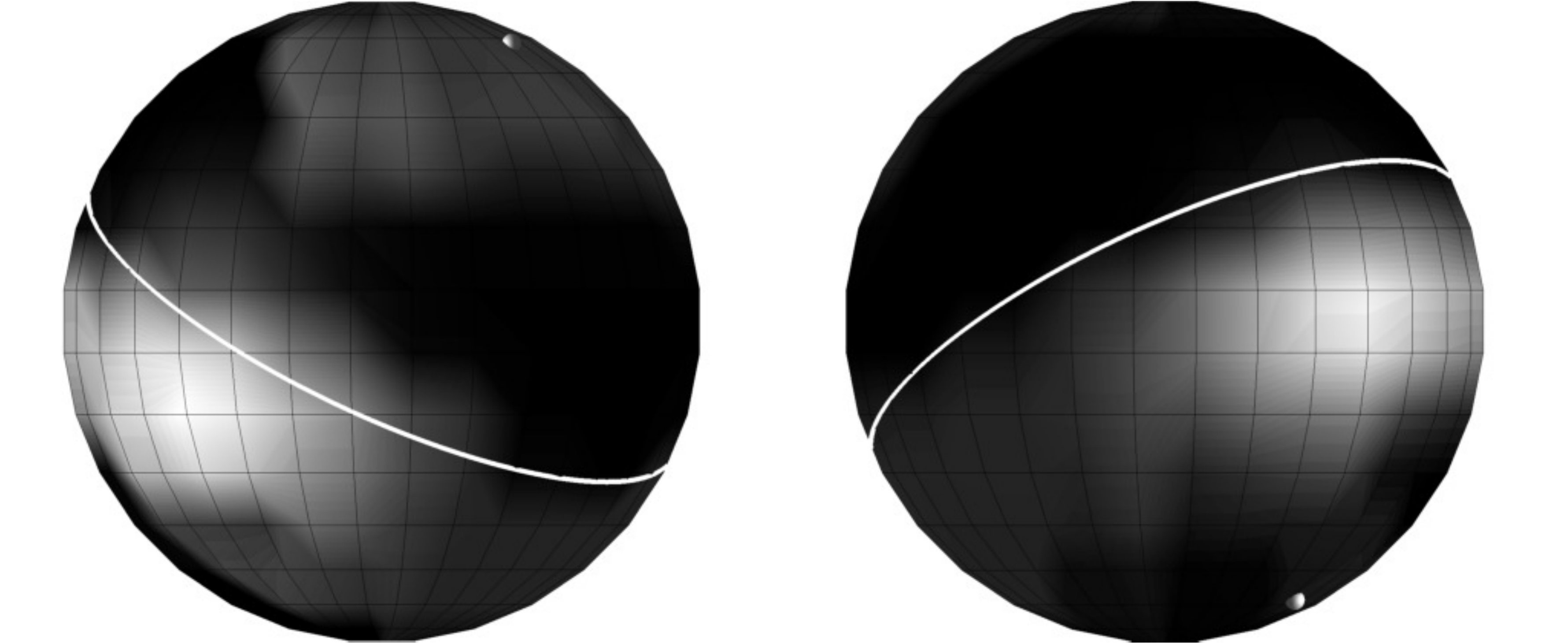}} \\
i) 0.8 & j) 0.9 \\
\end{tabular}
\caption{Distribution of the density common logarithm (in units of $\rho(L_1)$) from 
$-8$ to $-3$ on the accretor's surface for different phases of 
$P_{\text{beat}}$ period. In each diagram the left panel shows the western
hemisphere, the right one shows the eastern hemisphere (see explanations in 
the text concerning the orientation of the coordinate system). The illuminated 
side of the surface faces the donor star. The line denotes the magnetic 
equator. The balls point to locations of the northern (upper) and southern 
(lower) magnetic poles.}
\label{fg-spots}
\end{figure}

The distribution of the radiation originating on the accretor's surface at
different azimuthal angles (phases) is shown in Fig. \ref{fg-spots}, where the 
distributions of the common logarithm of the accreted matter density are 
shown. The diagrams correspond to phases from $0$ to $0.9$ of the 
$P_\text{beat}$ period of the model asynchronous polar. The order of the 
diagrams is from left to right and top-down. So that the two upper diagrams 
(a and b) correspond to phases $0$ (left, a) and $0.1$ (right, b). The next 
two diagrams (second row, c and d) correspond to phases $0.2$ (left, c) and 
$0.3$ (right, d) etc. Finally, the two lowest diagrams (i and j) display 
phases $0.8$ (left, i) and $0.9$ (right, j).

In all the diagrams the same gray scale is used. The limits of the scale are
$-8$ (minimal) and $-3$ (maximal). The dark (minimal density) regions 
correspond to the effective temperature of the white dwarf, taken to be
$T_\text{eff} = 37000$~K. The left panel of each diagram displays the
western hemisphere of the star and the right one shows the eastern hemisphere. 
The balls in the diagrams correspond to locations of the northern (upper) and
southern (lower) magnetic poles. The magnetic equator is denoted by the line.

\begin{figure}[t]
\centering
\includegraphics[width=0.95\textwidth]{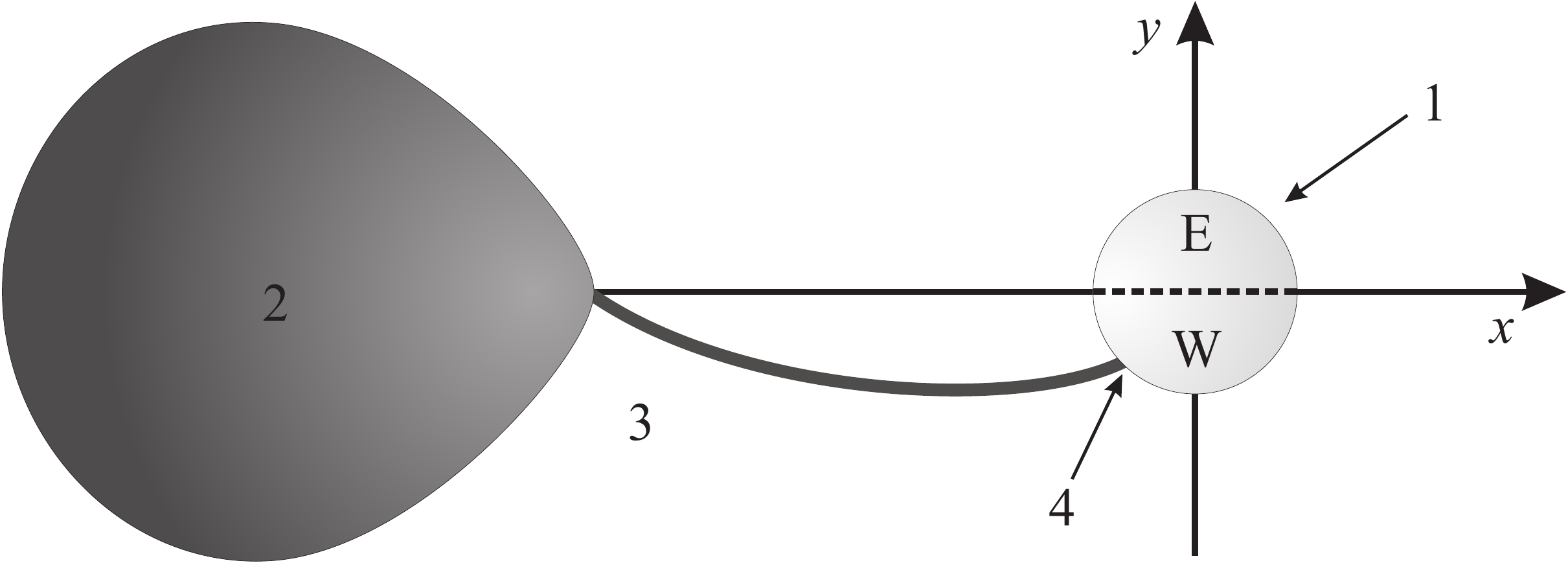}
\caption{The scheme explaining the definition of western (W) and eastern (E) 
hemispheres shown on Fig. \ref{fg-spots}. The numbers are corresponded to: 
1~--- accretor-star, 2~--- donor-star, 3~--- stream, 4~--- hot spot.}%
\label{fg-hemi}
\end{figure}

The images of the surface are oriented as follows (see Fig. \ref{fg-hemi}). 
The origin of the azimuthal coordinates (longitude) is at the point located on 
the opposite to the donor side of the accretor. We remind the reader that in 
our calculations we used a Cartesian coordinate system whose origin is in the 
accretor and the center of the donor is at the coordinate $x = -A$. The point 
of intersection of the zero meridian (right edge of the left hemisphere and 
left edge of the right hemisphere) and the equator lies in a point with the 
positive $x$ coordinate. This circumstance is emphasized in the diagrams shown 
by the distribution of illumination of the accretor's surface. The illuminated 
side faces the donor star and the shaded side is opposite.

Thus, in the diagrams, one can see density distributions on the accretor's
surface shown in a reference frame related to the donor. The coordinate frame
(meridians and parallel circles) on the accretor's surface does not describe
the true geographical coordinate system, since it is related to the donor but
not the accretor. In this coordinate system the accretor makes a full rotation
during the $P_\text{beat}$ period. So its true geographical coordinate system
always moves with respect to the drawn grid. Thus, everything looks as if we
had observed the accretor during the $P_\text{beat}$ period sitting in the 
$L_1$ point or any other point resting in the coordinate system that rotates 
with the binary system.

Our analysis shows that the geometry of the hot spots on the accretor's 
surface is rather complex. However it exactly corresponds to the description 
of the 3-D flow structure given in the previous section. All the diagrams 
demonstrate two main zones of accretion: in the vicinity of the northern 
magnetic pole and along the magnetic belt located a little lower than the 
magnetic equator. At the initial phase (upper left diagram, a) the main hot 
spot is located in the western hemisphere. On the opposite side in the eastern 
hemisphere one can see an analogous hot spot but of lower intensity. Formation 
of these two spots is in exact agreement with the accretion pattern shown in 
Fig. \ref{fg-scheme}. The third spot that is also of low intensity forms on 
the northern magnetic pole. The similar distribution of the hot spots is 
observed at phases $0.1$, $0.2$ and $0.3$ (diagrams b, c and d).

As we noted above at phase $0.4$ the flow structure changes. The main 
accretion flow splits into two flows. As a result the hot spot near the 
magnetic belt becomes weaker and, oppositely, the hot spot on the northern 
magnetic pole becomes more intense. At phase $0.5$ (diagram f) we can see
two corresponding spots of approximately the same intensity. At phases $0.6$,
$0.7$ and $0.8$ the main hot spot is on the northern magnetic pole. The second
spot near the magnetic belt disappears. Finally, at the phase $0.9$ (diagram 
j) the accretion stream splits again and forms two simultaneous hot spots.

Thus during the $P_\text{beat}$ period the intensity, location and number of 
the hot spots varies in a non-trivial manner. Equivalently, we see 
considerable variation in our sample of ten polars as a function of azimuthal 
angle. We should note that the radiation originating from the magnetic pole is 
significantly polarized since the magnetic field in this region has a 
preferential direction. At the same time the radiation of the hot spot located 
near the magnetic belt is not polarized since the magnetic field in this 
region has no preferential directions.

\begin{figure}[t]
\centering
\includegraphics[width=0.95\textwidth]{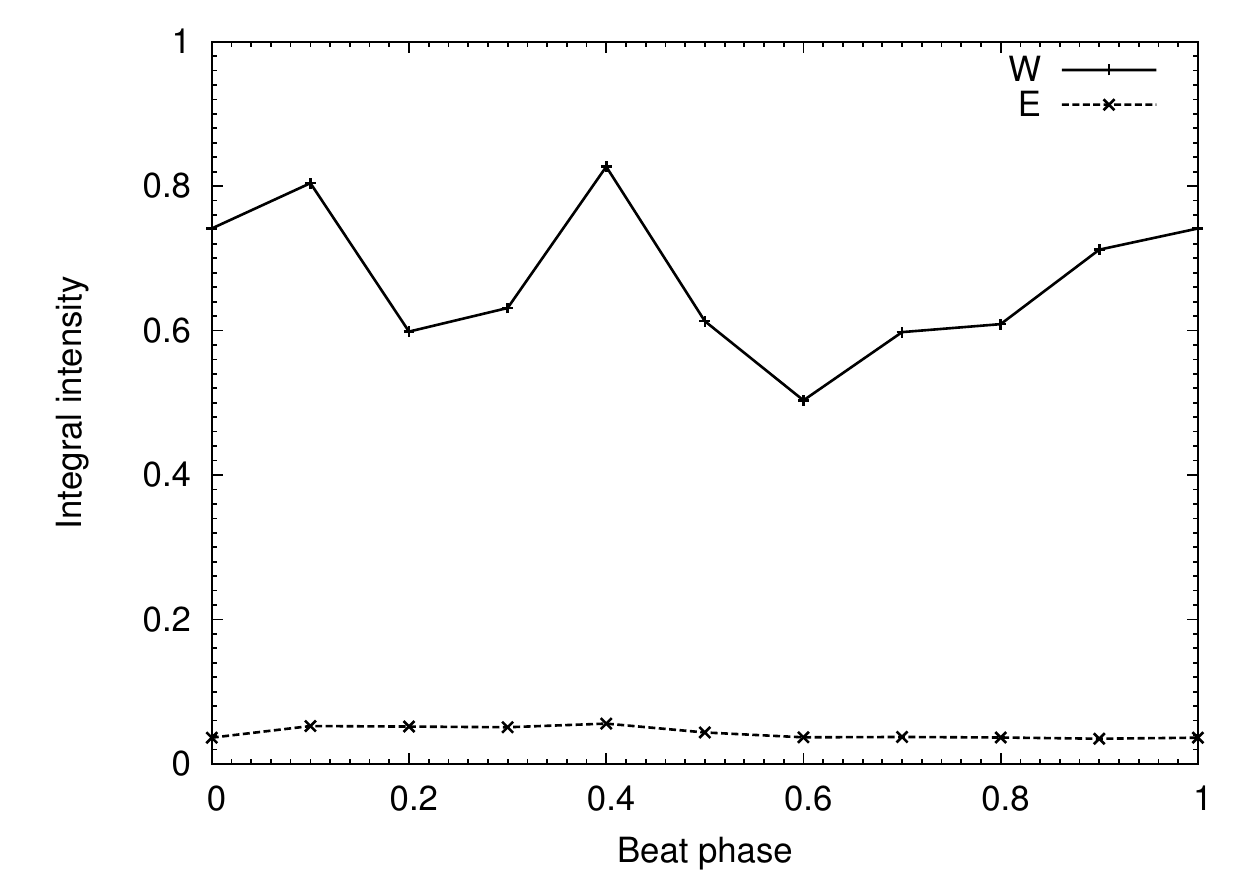}%
\caption{Variations of the total intensity of the radiation coming from the western 
and eastern hemispheres of the accretor's surface during the $P_\text{beat}$ 
period.}%
\label{fg-intensity}
\end{figure}

In Figure \ref{fg-intensity} we show curves describing variations of the total
intensity of the radiation (in arbitrary units) coming from the accretor's
surface during the $P_\text{beat}$ period. The solid curve corresponds to the
western hemisphere and the dashed line corresponds to the eastern hemisphere.
Figure shows that the radiation that of the western hemisphere is much more
intense than of the eastern hemisphere. This is because the accretion stream
approaches the accretor from the side of the western hemisphere due to the
rotation of the binary. Thus most of the energy is released exactly in this
region.

In the curve corresponding to the western hemisphere one can see two maxima 
and two minima. So, during the $P_\text{beat}$ period we see two bursts of
intensity. The amplitude of the intensity variations can reach 65\%. The first
burst starts from the deep minimum at the phase $0.6$ and finishes at the 
phase $0.1$. The second burst is more sharp and approaches its maximum at the 
phase $0.4$. It is interesting to note that at these phases ($0.1$ and $0.4$) 
the maxima are observed also in the curve corresponding to the eastern 
hemisphere. The intensity of the radiation coming from the eastern hemisphere 
varies with an amplitude of up to 60\%.

Comparison of the curves in Fig. \ref{fg-intensity} with the diagrams of Fig.
\ref{fg-spots} shows that the intensity maxima occur at time moments when the
hot spot near the magnetic belt is located exactly on the geographical equator
of the accretor. The radiation flux from the second hot spot located on the
northern magnetic pole is weaker since it is on the edge of the accretor's
limb.

\subsection{Synthetic light curves}

To make the presentation of the simulation results and comparison of them with
observations more illustrative we computed synthetic light curves. They allow
one to trace and investigate variations of the intensity of the systems's
radiation concerned with variations of the location and number of the hot 
spots on the accretor's surface. To compute the light curves we used the
method described by Romanova et al. \cite{Romanova2004}.

The intensity of the radiation originating on the accretor's surface was
calculated using the following relation:
\begin{equation}\label{eq15}
 I = \dfrac{1}{\pi} \int\limits_{\cos\beta > 0} f(\vec{R}) \cos\beta\, dS,
\end{equation}
where $dS$ is the element of the spherical surface, $\cos\beta = \vec{k}
\cdot \vec{n}$, $\vec{k}$ is the unit vector directed from the given point
of the surface to the observer. The vector $\vec{k}$ is determined by the
angle $i$ of the inclination of the orbital plane with respect to the picture
plane: $\cos i = \vec{\Omega} \cdot \vec{k}/\Omega$. The integration in
\eqref{eq15} is performed over only the visible for the observer side of the
hemisphere ($\cos\beta > 0$). The absence of eclipses in this model means that 
the range of allowed inclination angles must be $i \lesssim 60^\circ$ or $i 
\gtrsim 120^\circ$.

\begin{figure}[p]
\centering
\begin{tabular}{cc}
\hbox{\includegraphics[width=0.3\textwidth]{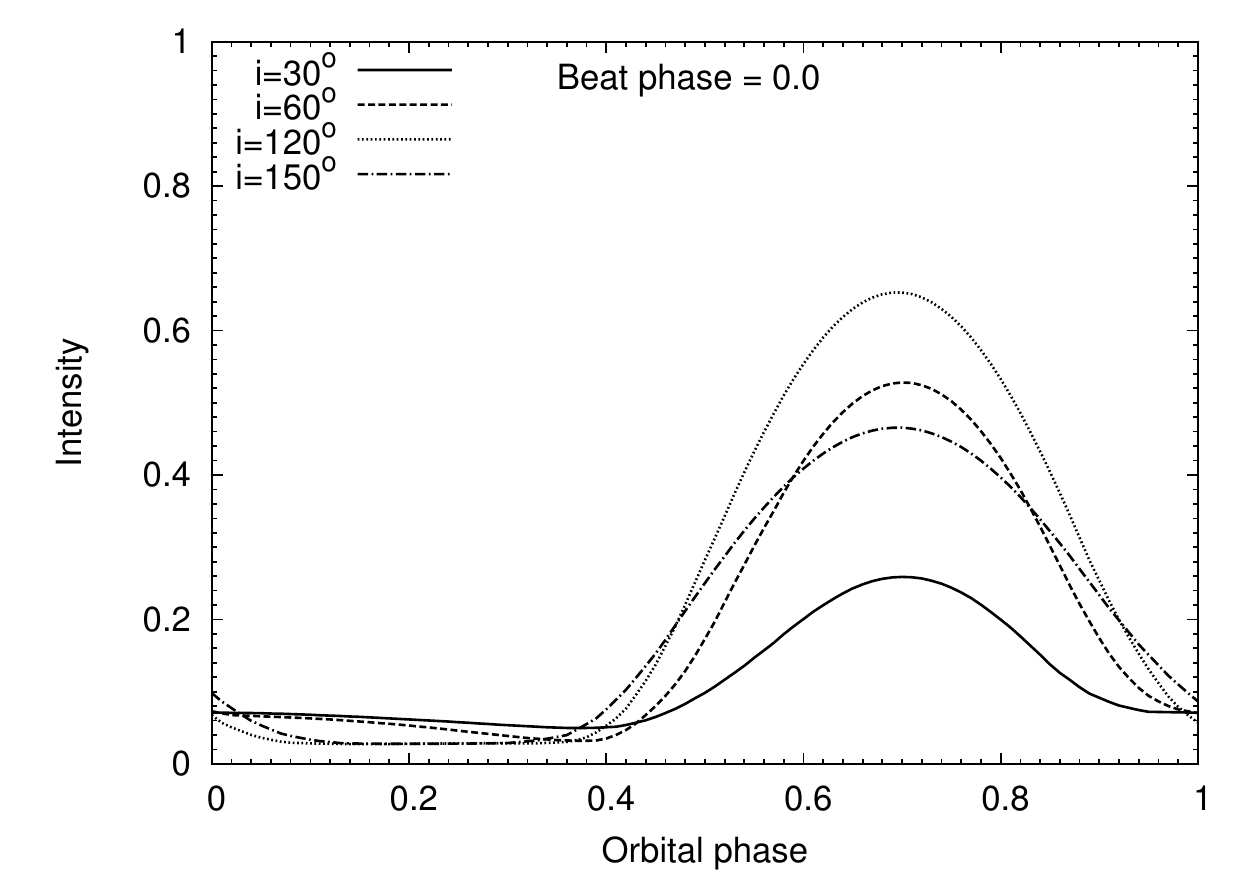}} &
\hbox{\includegraphics[width=0.3\textwidth]{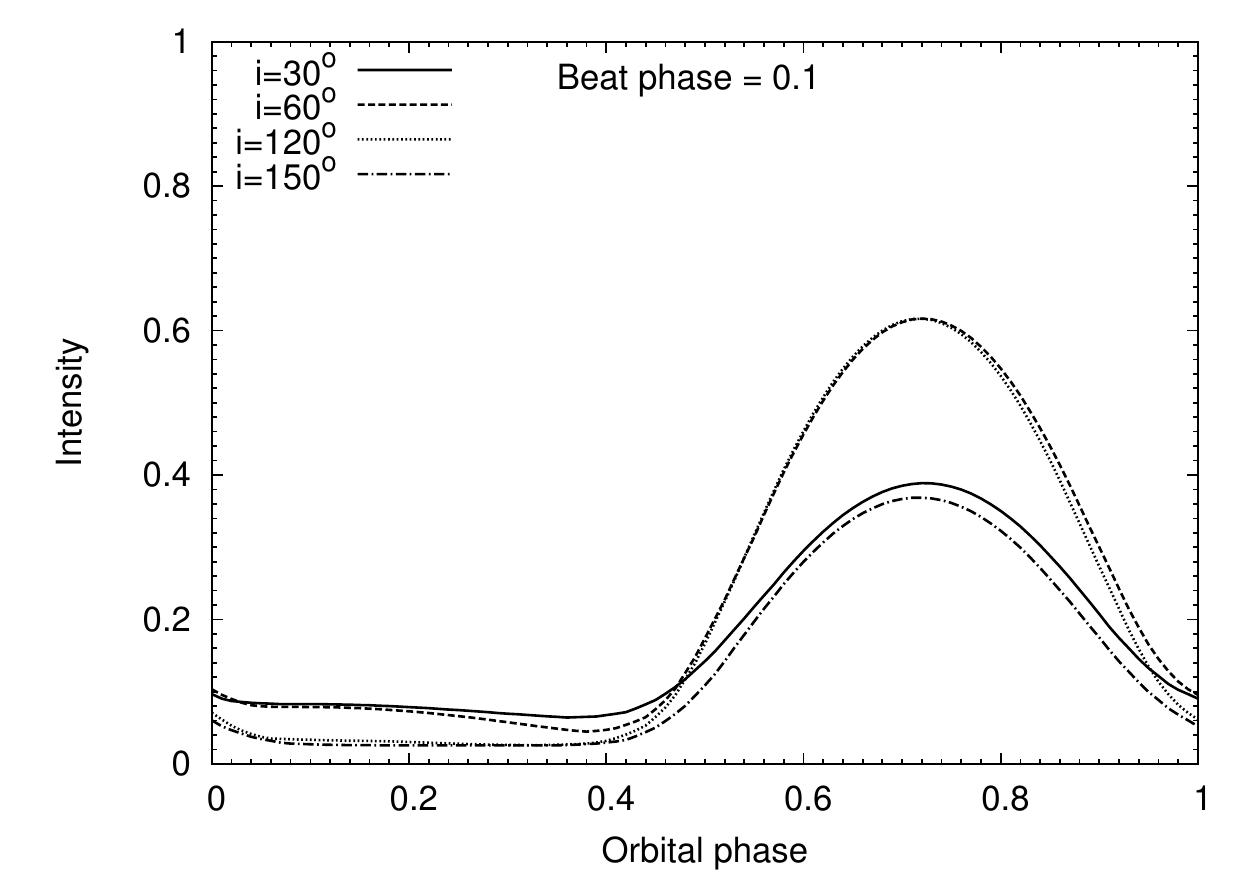}} \\ 
\hbox{\includegraphics[width=0.3\textwidth]{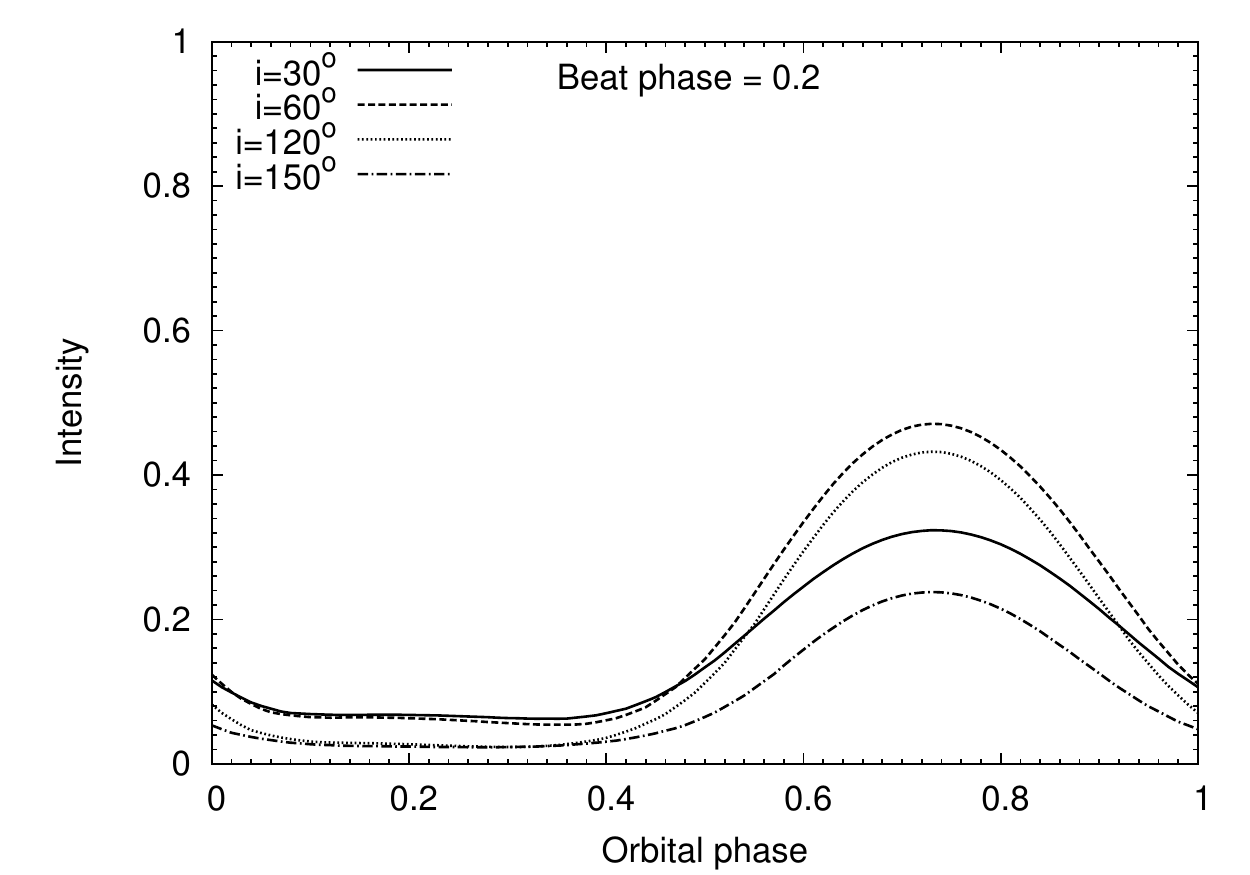}} &
\hbox{\includegraphics[width=0.3\textwidth]{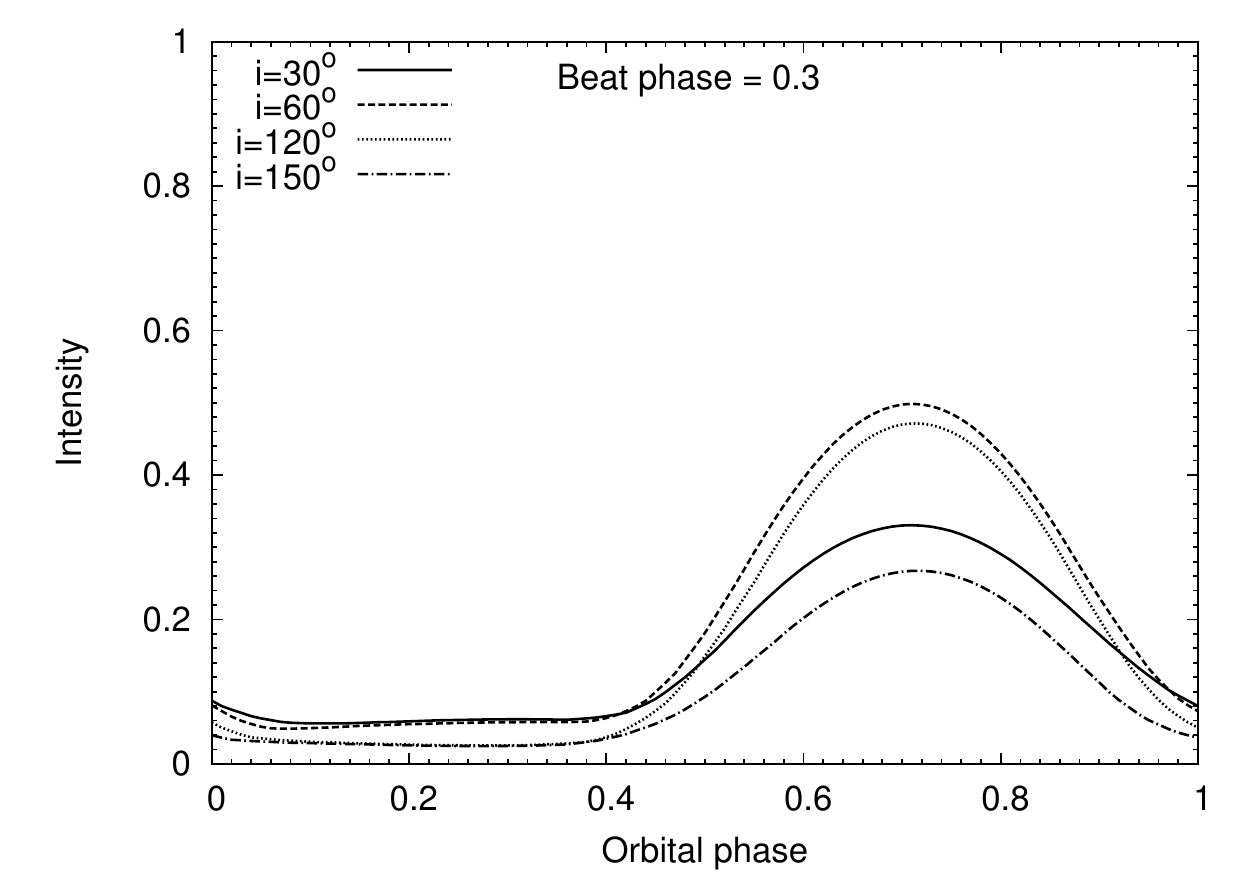}} \\ 
\hbox{\includegraphics[width=0.3\textwidth]{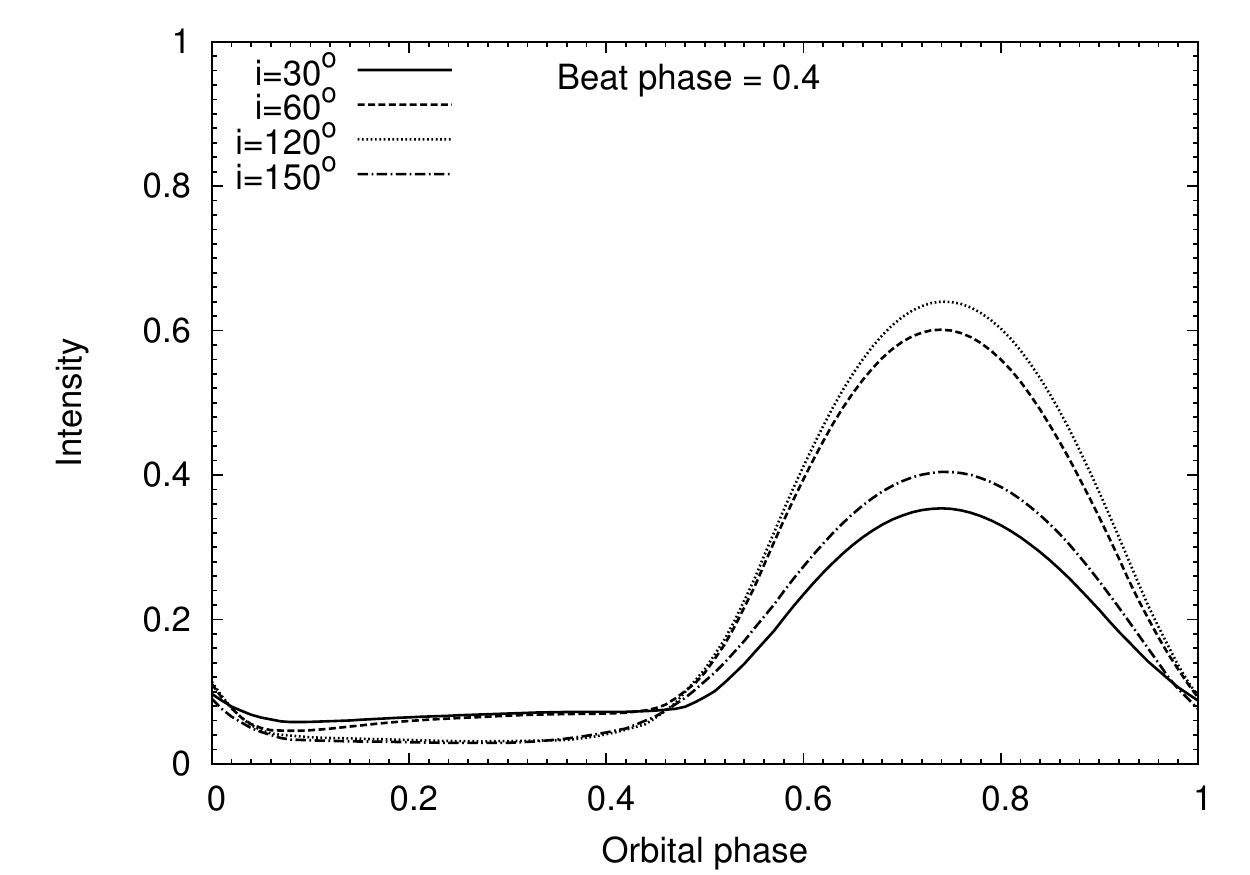}} &
\hbox{\includegraphics[width=0.3\textwidth]{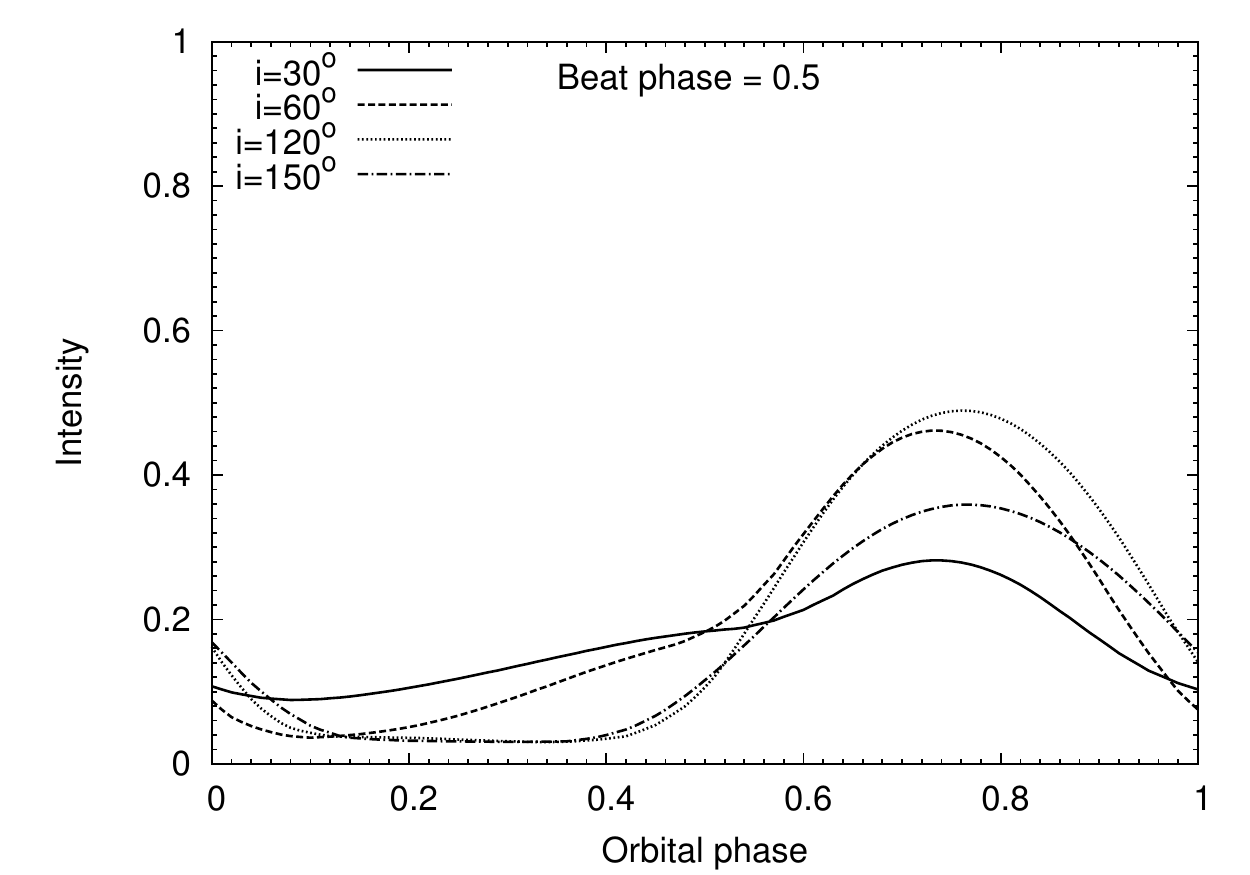}} \\ 
\hbox{\includegraphics[width=0.3\textwidth]{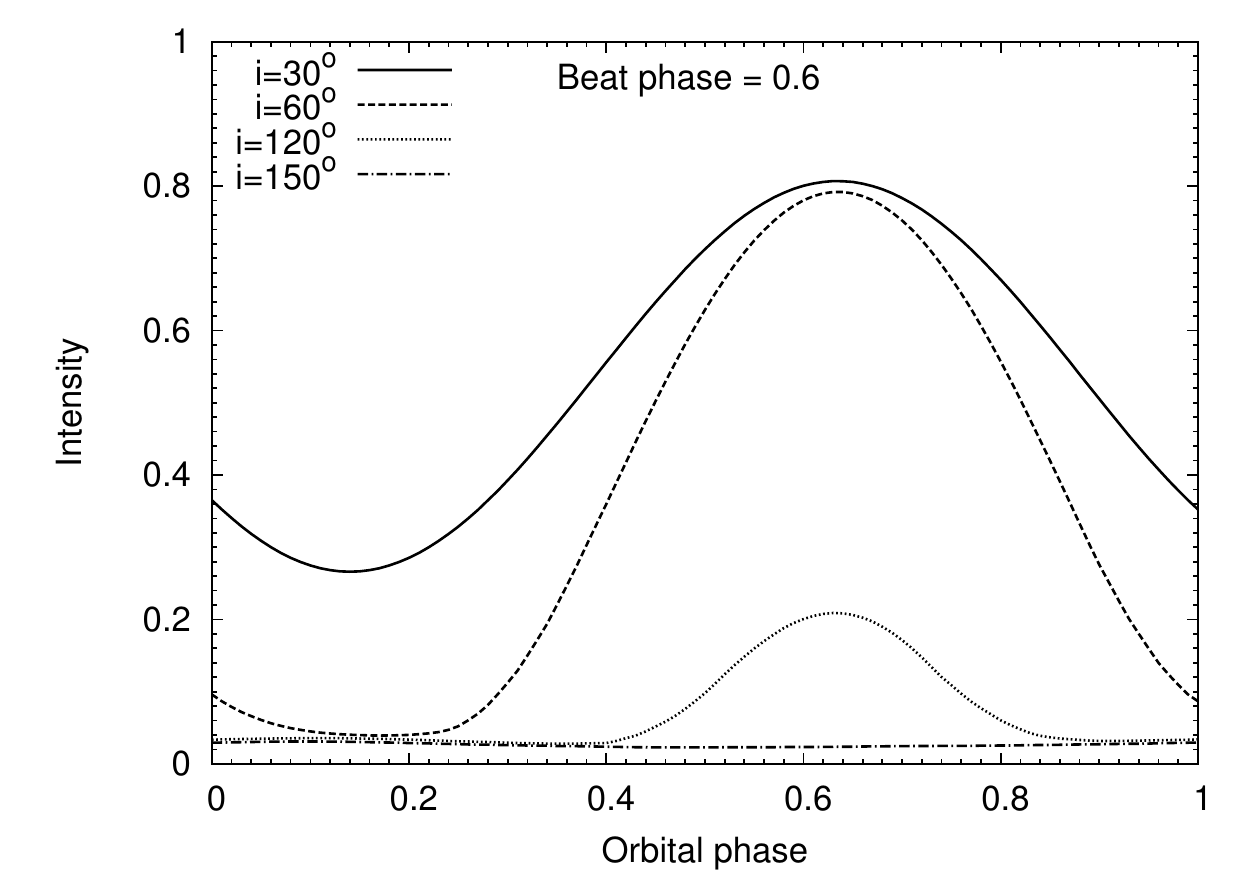}} &
\hbox{\includegraphics[width=0.3\textwidth]{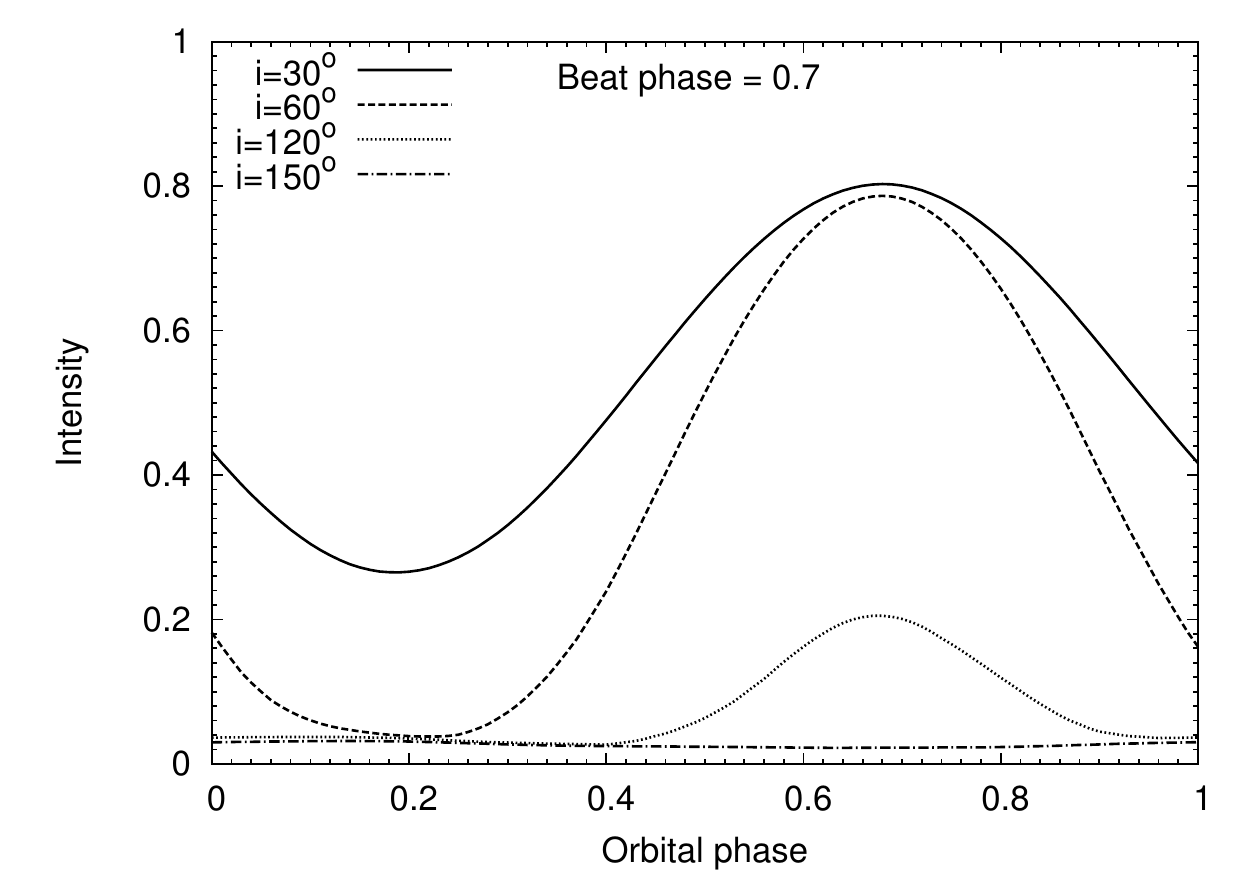}} \\ 
\hbox{\includegraphics[width=0.3\textwidth]{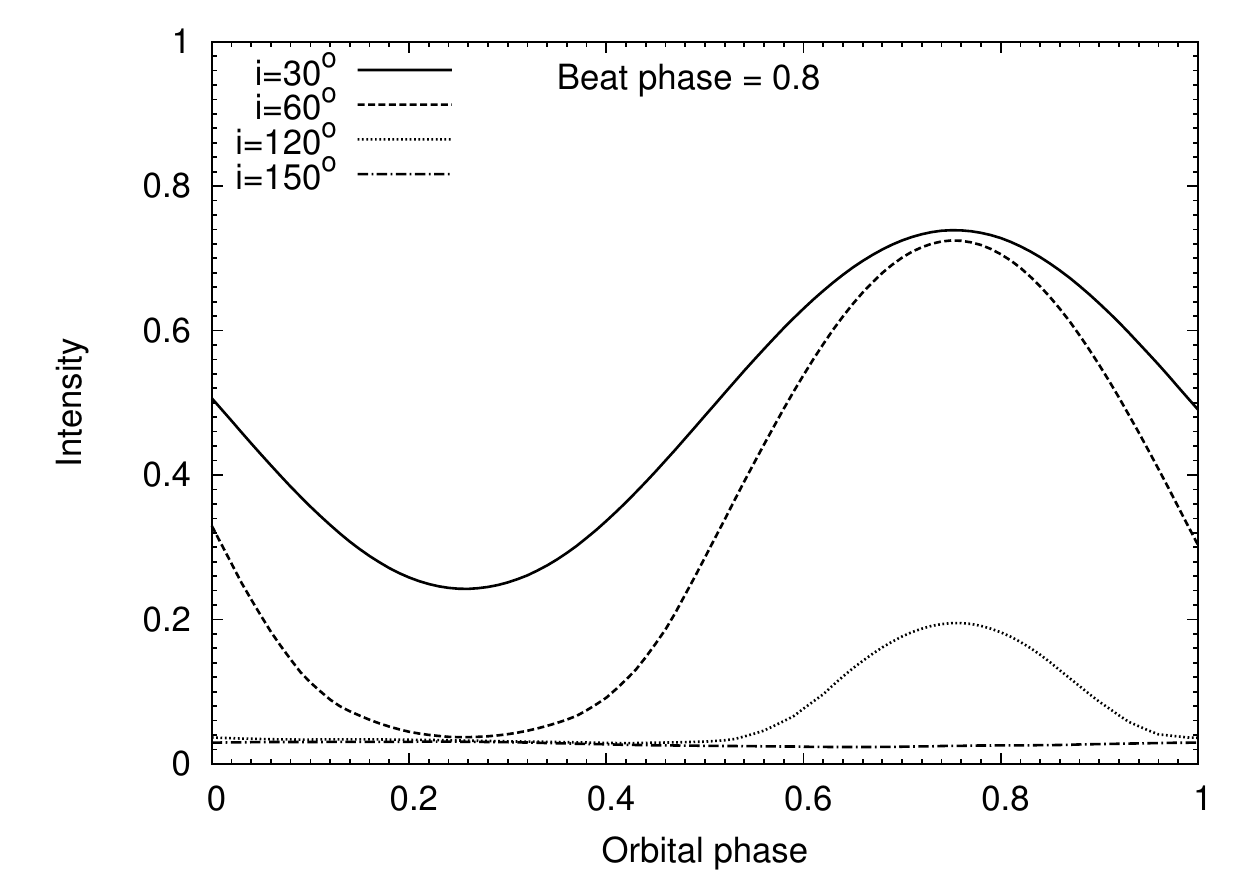}} &
\hbox{\includegraphics[width=0.3\textwidth]{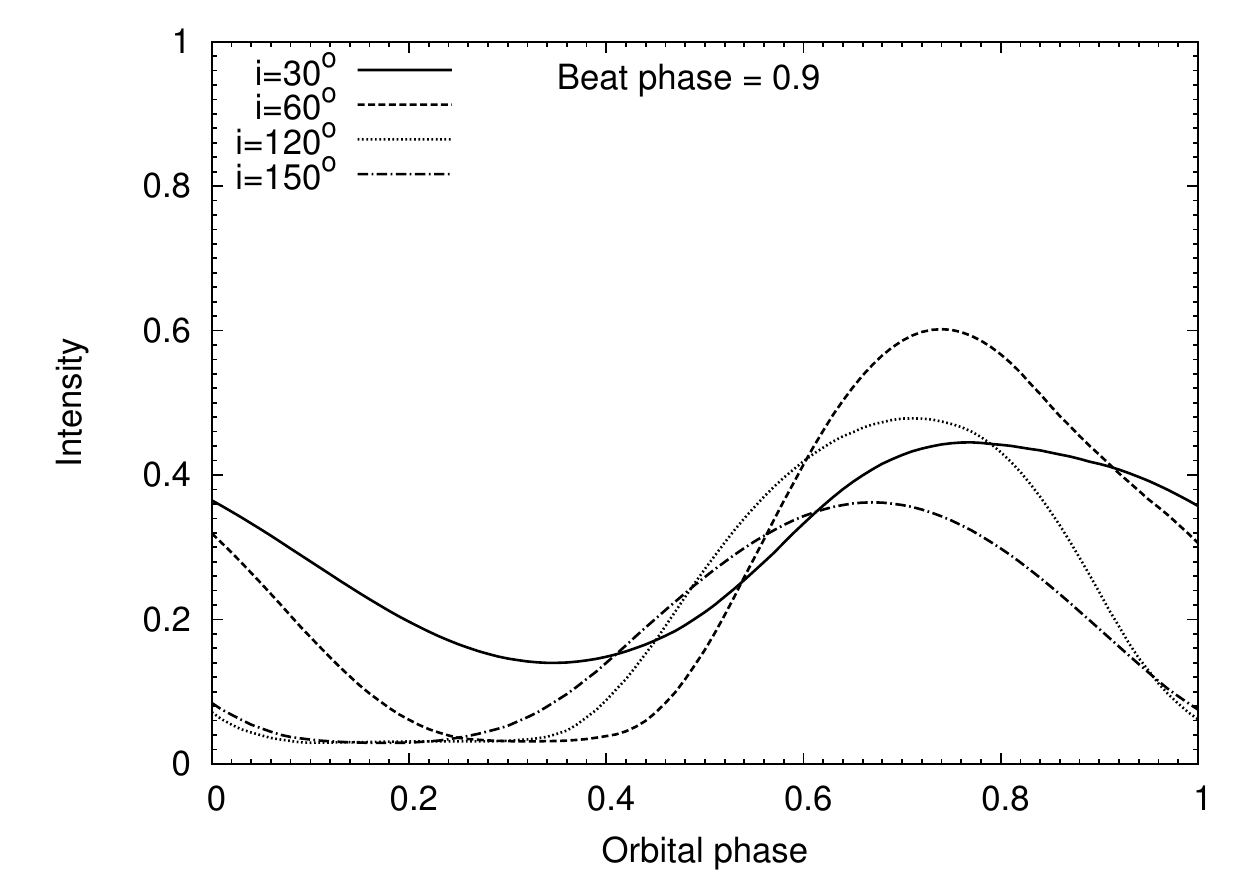}} \\ 
\end{tabular}
\caption{Synthetic light curves computed for different phases of the $P_\text{beat}$ 
period. The different curves in the graphs correspond to different inclination 
angles $i$.
}
\label{fg-lc}
\end{figure}

The resulting synthetic light curves are shown in Fig. \ref{fg-lc}. The graphs
correspond to different azimuthal angles or phases of the $P_\text{beat}$ 
period. The different curves in the graphs correspond to different inclination 
angles $i$ of the binary.

Analysis of the light curves show that during the $P_\text{beat}$ period the
amplitude of the radiation intensity may vary up to factor of $2$ depending on
the value of the inclination, $i$. The first half of the $P_\text{beat}$ 
period is relatively quiescent since the accretor in this period faces the 
observer with its eastern hemisphere. Since, from the analysis there are no 
intense hot spots formed in the eastern hemisphere at those times. The second 
half of the $P_\text{beat}$ period is, in contrast, much more active. During 
these beat phases, the accretor points its western hemisphere to the 
secondary. The hot spots which are formed at the foot-points of the accretion 
stream in the western hemisphere cause an increase of the system brightness at 
these beat phases. Observed light curves of BY Cam \cite{Silber1997, 
Mason1998, Pavlenko2007a, Andronov2008} demonstrate very similar behaviour, as 
they also very by a factor $\sim$ 2 as a function of spin-orbit beat phase.

Another interesting effect observed in the synthetic light curves is the 
result of shifts in positions of minima and maxima during the $P_\text{beat}$ 
period. For example, at phase $0.6$ the brightness maximum is achieved at the 
orbital phase $0.6$ while at spin-orbit beat phases of $0.8$ and $0.9$ it is 
achieved at the orbital phase $0.8$. The brightness minimum at the spin-orbit 
phase $0.6$ corresponds to the orbital phase $0.1$ while at spin-orbit phase 
$0.9$ this minimum occurs at orbital phase $0.4$. We note that similar phase 
shifting occurs in the observed light curves of BY Cam \cite{Silber1997, 
Mason1998, Pavlenko2007a, Andronov2008}. 

This sort of phase shifting is strong circumstantial evidence for the 
existence of an equatorial accretion region in BY Cam, as the dipolar 
accretion spot does not result in significant phase shifting as a function of 
azimuthal angle. The results of our calculations are encouraging, but we 
expect that further exploration of polar parameter space may yield significant 
refinement of this first attempt at asynchronous polar modelling with complex 
magnetic fields. 

\section{Conclusions}

In the paper we present a numerical study designed to investigate the mass
transfer process in cataclysmic variables where a strong magnetic field having 
complex geometry exists. In our model, we suppose that the magnetic field of 
the accretor is a superposition of aligned dipole and quadrupole field 
components. The model is based on the following assumption. Namely, that 
plasma dynamics in the accretion stream issuing from the donor's envelope 
through the inner Lagrangian point $L_1$ are determined by the slow average 
motion of matter and very rapid MHD waves propagating against a background of 
this slow motion. The strong background magnetic field plays the role of an 
effective fluid interacting with plasma. In the framework of this modified 
MHD, it is possible to simulate flows in close binary systems where the 
magnetic field on the accretor's surface reaches values of $10^7$--$10^8$~G, 
corresponding to the case of polars. In the numerical model we also have 
taken into account processes of the diffusion of the magnetic field due to 
the MHD wave turbulence and processes of radiative heating and cooling.

Using a magnetic field model developed and implemented as a modified 3-D 
parallel numerical code, we performed simulations of the MHD flow structure in
an ensemble of 10 synchronized polars. We present results of the calculations 
for accretion flows for ten polar orientations, with only azimuthal angle 
variations. The results are also applied to asynchronous polars and determined 
to be valid as long as the spin and orbital periods differ by at most a few 
percent. In the framework of the asynchronous polar model these azimuthal 
angles correspond to ten different phases around the $P_\text{beat}$ period 
that is determined by the period of the proper rotation of the accretor in 
the reference frame related to the binary system. 

The basic conclusions of our work are as follows. 

1. In close binary systems where the accretor's field is strong and purely 
dipolar, the accretion of matter (with formation of corresponding hot spots) 
may impact either onto the northern magnetic pole, or the southern pole or 
both \cite{ZBSMF2010}. From an ensemble of ten synchronous polars with complex 
magnetic fields,  it is shown that the existence of a strong quadrupole 
component of the magnetic field leads to significant complications of the 
accretion flow. The affect on the accretion flow and resulting accretion 
impact zones are strong functions of azimuthal angle. We find that synthetic 
light curves vary in shape and overall intensity as a function of azimuthal 
angle due to the visibility of the resulting accretion regions.

2. Accretion modes in polars with complex fields include both single-pole and 
two-pole configurations as in the pure dipole case. However, in the complex 
field case, the single-pole polar may have either a spot on the magnetic pole 
or alternatively, it could have an accretion region on the magnetic equator, 
near the orbital plane. In such cases, the observations may be mistakenly 
described as purely dipole magnetic field. If a strong quadrupole component of 
the field is present then additional accretion zones and corresponding hot 
spots form near the magnetic belt located a little bit south of the magnetic 
equator. The radiation from hot spots located on the magnetic poles is 
strongly polarized; since, in these zones the magnetic field has a 
preferential direction. At the same time the radiation of the hot spots 
located near the magnetic belt does not exhibit polarization since in this 
region no preferential direction of the field exists.

3. Analysis of the light curves computed in this work shows that at orbital 
phases higher than $0.5$ the intensity of the radiation sharply increases (by 
factor of 2 and more). This is because at these orbital phases the main 
accretion hot spots face the observer. In addition, positions of maxima and 
minima of the light curve shift during the $P_\text{beat}$ period of our model 
asynchronous polar. Similar non-trivial behaviour is also shown by observed 
light curves of the BY Cam system \cite{Silber1997, Mason1998, Pavlenko2007a}.
However, in BY Cam the working model involves pole switching between two 
equatorial poles, while in the present model pole switching occurs between a 
polar spot and an equatorial spot. More studies are needed to determine if 
either of these complex field models can be confidently applied to BY Cam.

There is a vast parameter space, yet to be explored, to study magnetic stream
accretion. Parameters that may be varied include, the inclination angle of the 
magnetic axis, the ratio of intensities of the dipole and quadrupole 
components of the field, the accretion rate, mass-ratio etc. However, in this 
work we did not aim to achieve the exact correspondence of the calculation 
results and observations. We focused our attention on the main properties of 
physics of the mass transfer in complex field mCVs and possible observational 
signatures of complex field geometry. More precise adjustment of model 
parameters are expected to achieve correspondence with observational data on 
particular mCVs, especially asynchronous polars. Such studies are needed in 
order to make further progress in understanding accretion in the presence of 
complex magnetic fields. We conclusively find that 3-D MHD calculations 
simulating a stream accreting magnetic white dwarf can be used to properly 
model complex magnetic field structure in mCVs.\\

This work was supported by the Basic Research Program of the Presidium of the 
Russian Academy of Sciences, Russian Foundation for Basic Research (projects 
09-02-00064, 11-02-00076), Federal Targeted Program ''Science and Science 
Education for Innovation in Russia 2009-2013''.

\end{document}